\documentclass[12pt]{iopart}

\expandafter\let\csname equation*\endcsname\relax
\expandafter\let\csname endequation*\endcsname\relax

\usepackage{graphicx} 
\usepackage{amsmath}
\usepackage{amssymb}
\usepackage{color}
\usepackage{cite}

\def \Nipyz {[Ni(H$_2$O)$_2$(pyrazine)$_2$](BF$_4$)$_2$}
\def \Nilut {[Ni(H$_2$O)$_2$(3,5-lutidine)$_4$](BF$_4$)$_2$}
\def \Niace {Ni(H$_2$O)$_2$(acetate)$_2$(4-picoline)$_2$}

\begin{document}
\title[Determining parameters of polycrystalline spin-1
  magnets]{Determining the anisotropy and exchange parameters of
  polycrystalline spin-1 magnets}
%\title{Extracting magnetic information from powdered spin-1 quantum antiferromagnets}

\author{W. J. A. Blackmore$^1$,
J. Brambleby$^1$,
T. Lancaster$^{2}$,
S. J. Clark$^{2}$,
R. D. Johnson$^3$,
J. Singleton$^4$,
A.~Ozarowski$^5$,
J. A. Schlueter$^{6,7}$,
Y.-S. Chen$^8$,
A. M. Arif\,$^9$,
S. Lapidus$^{10}$,
F. Xiao$^{11,12}$,
R. C. Williams$^1$, 
S. J. Blundell$^{3}$,
M. J. Pearce$^1$,
M. R. Lees $^1$,
P. Manuel$^{13}$,
D. Y. Villa$^{14}$,
J. A. Villa$^{14}$,
J. L. Manson$^{14}$ and
P. A. Goddard$^1$}

\address{$^1$Department of Physics, University of Warwick, Coventry, CV4 7AL, UK}
\address{$^{2}$Centre for Materials Physics, Durham University, Durham, DH1 3LE, UK}
\address{$^{3}$Clarendon Laboratory, University of Oxford, Parks Road, Oxford OX1 3PU, UK}
\address{$^4$NHMFL, Los Alamos National Laboratory, Los Alamos, NM 87545, USA}
\address{$^5$NHMFL, Florida State University, Tallahassee, FL 32310, USA}
\address{$^6$Materials Science Division, Argonne National Laboratory, IL 60439, USA}
\address{$^7$Division of Materials Research, National Science Foundation, 2415 Eisenhower Ave, Alexandria, VA 22314, USA}
\address{$^8$ChemMatCARS, Advanced Photon Source, Argonne National Laboratory, IL 60439, USA}
\address{$^9$Department of Chemistry, University of Utah, Salt Lake City, UT 84112, USA}
\address{$^{10}$X-ray Sciences Division, Argonne National Laboratory, IL 60439, USA}
\address{$^{11}$Department of Chemistry and Biochemistry, University of Bern, 3012 Bern, Switzerland}
\address{$^{12}$Laboratory for Neutron Scattering and Imaging, Paul Scherrer Institut, CH-5232 Villigen PSI, Switzerland}
\address{$^{13}$ISIS Facility, Rutherford Appleton Laboratory, Chilton, Didcot, Oxon, OX11 0QX, UK}
\address{$^{14}$Department of Chemistry and Biochemistry, Eastern Washington State University, Cheney, WA 99004, USA}
\eads{\mailto{jmanson@ewu.edu}, \mailto{p.goddard@warwick.ac.uk}}

\begin{abstract}
Although low-dimensional $S = 1$ antiferromagnets remain of great
interest, difficulty in obtaining high-quality single crystals of
the newest materials hinders experimental research in this
area. Polycrystalline samples are more readily produced, but there are
inherent problems in extracting the magnetic properties of anisotropic
systems from powder data. Following a discussion of the effect of
powder-averaging on various measurement techniques, we present a
methodology to overcome this issue using thermodynamic
measurements. In particular we focus on whether it is possible to
characterise the magnetic properties of polycrystalline, anisotropic
samples using readily available laboratory equipment. We test the
efficacy of our method using the magnets \Nilut\ and
\Niace, which have negligible exchange interactions, as well as the antiferromagnet \Nipyz, and show that we are able to
extract the anisotropy parameters in each case. The results obtained
from the thermodynamic measurements are checked against electron-spin
resonance and neutron diffraction. We also present a
 density functional method, which incorporates spin-orbit coupling to
 estimate the size of the anisotropy in \Nipyz.
\end{abstract}

\section{Introduction}

The investigation of low-dimensional quantum magnets is a key thrust
of condensed matter physics. Of particular relevance here are
quasi-one-dimensional and quasi-two-dimensional systems based on $S =
1$ magnetic moments, which inspire a great deal of contemporary
theoretical attention
(e.g.~\cite{intro1,intro2,intro3,intro4,intro5,intro6,intro7,intro8})
and are predicted to display vibrant phase diagrams arising from
competing interactions and their interplay with single-ion
anisotropy. These diagrams encompass quantum critical points~\cite{intro1,intro2}, nematic and supersolid states~\cite{intro5,keola3,keola4}, 
as well as topologically interesting gapped and quantum paramagnetic phases~\cite{haldane,aklt,keola1,keola2}. 
By contrast, because of the difficulty in making real examples of these
systems, experimental work in this area moves more slowly and several
predictions remain untested. While recent advances have been made with
molecule-based
magnets~\cite{filho04,zapf06,yin08,NiSbF6,manson12,chen16,Junjie,JamieBSbF6,kagome},
difficulties in obtaining high-quality single crystals of the newest
materials continue to hinder progress.

Detailed thermodynamic studies of single-crystal samples can be used
to find anisotropy parameters and the size of the primary magnetic
interactions, and reveal the ground state of the system. Unfortunately,
crystals of sufficient size for such measurements are often hard to
come by, particularly in the case of the newest materials, which are frequently
synthesized initially as powders. Optimising the synthesis procedures
for clean crystallization of a particular material typically requires
considerable time and effort. It is therefore advantageous to be able
to characterise the basic properties of a powdered anisotropic
magnetic material using simple, readily accessible measurement
techniques in order to identify the compounds that merit the additional work
required for crystal growth. However, the complication of
powder-averaging leads to difficulties interpreting the results of
bulk thermodynamic measurements. This issue is made worse if the
magnitude of the anisotropy is on a similar energy scale to the
strength of exchange interactions in the
compound~\cite{herchel07,NiSbF6, JamieBSbF6}.

Here we discuss the interpretation of experiments on polycrystalline
samples of $S = 1$ magnets under the influence of uniaxial and
rhombic single-ion anisotropy with energy $D$ and $E$,
respectively, and with the possibility of nearest-neighbour Heisenberg
exchange $J$ acting between spins.  We describe the effect of powder
averaging and discuss to what extent the parameters in the
Hamiltonian can be extracted from data, focusing particularly on bulk
thermodynamic measurements of susceptibility, magnetization and heat
capacity that can, in principle, be performed using commonly available
laboratory apparatus without the need to access equipment at a large
user facility. We will start by showing that for materials with anisotropic spins, but  
negligible exchange interactions, a good estimation of the
parameters can be readily extracted from powder data for both
easy-plane and easy-axis systems, provided low enough temperatures and
high enough magnetic fields can be achieved. We next consider the more
challenging situation in which antiferromagnetic exchange interactions
are finite and similar in energy to the single-ion anisotropy. We test the 
reliability of our methods by comparing the findings
derived from thermodynamic probes with additional facility measurements of
neutron diffraction and high-frequency electron-spin resonance
(ESR). Finally we describe an approach, using density
 functional theory and spin-orbit coupling, to provide reliable
estimates of single-ion anisotropy. 

We will apply the analysis to three new
materials in which Ni(II) ions are separated from one another by
molecular ligands. These are (1) \Nilut\, and (2) \Niace, both of which are found to be dominated by single-ion anisotropy with no evidence of a significant role for exchange interactions at the temperatures measured, and (3) the antiferromagnet \Nipyz. System (1) was designed to have NiN$_4$O$_2$ octahedra similar to that of (3), but without extended interaction pathways, such that the effect of the local environment on the anisotropy could be elucidated in the absence of exchange, and we will discuss to what extent this approach has been successful. 

While a subset of the methods outlined have been preliminarily tested in studies by some of the same authors~\cite{JamieBSbF6,kagome}, we combine the full methodology here for the first time. This work follows from a related investigation of how to extract exchange parameters in low-dimensional $S=1/2$ antiferromagnets~\cite{pagnjp}. 

The results presented here are of relevance not only to the study of low-dimensional $S=1$ magnets, but also to the growing field of Ni(II) single-ion magnets~\cite{frost16,miklovic15,marriot15,gruden14,ruamps13}.

\section{Systems with negligible exchange}

\begin{figure}[t]
\centering  
\includegraphics[width=10.0cm]{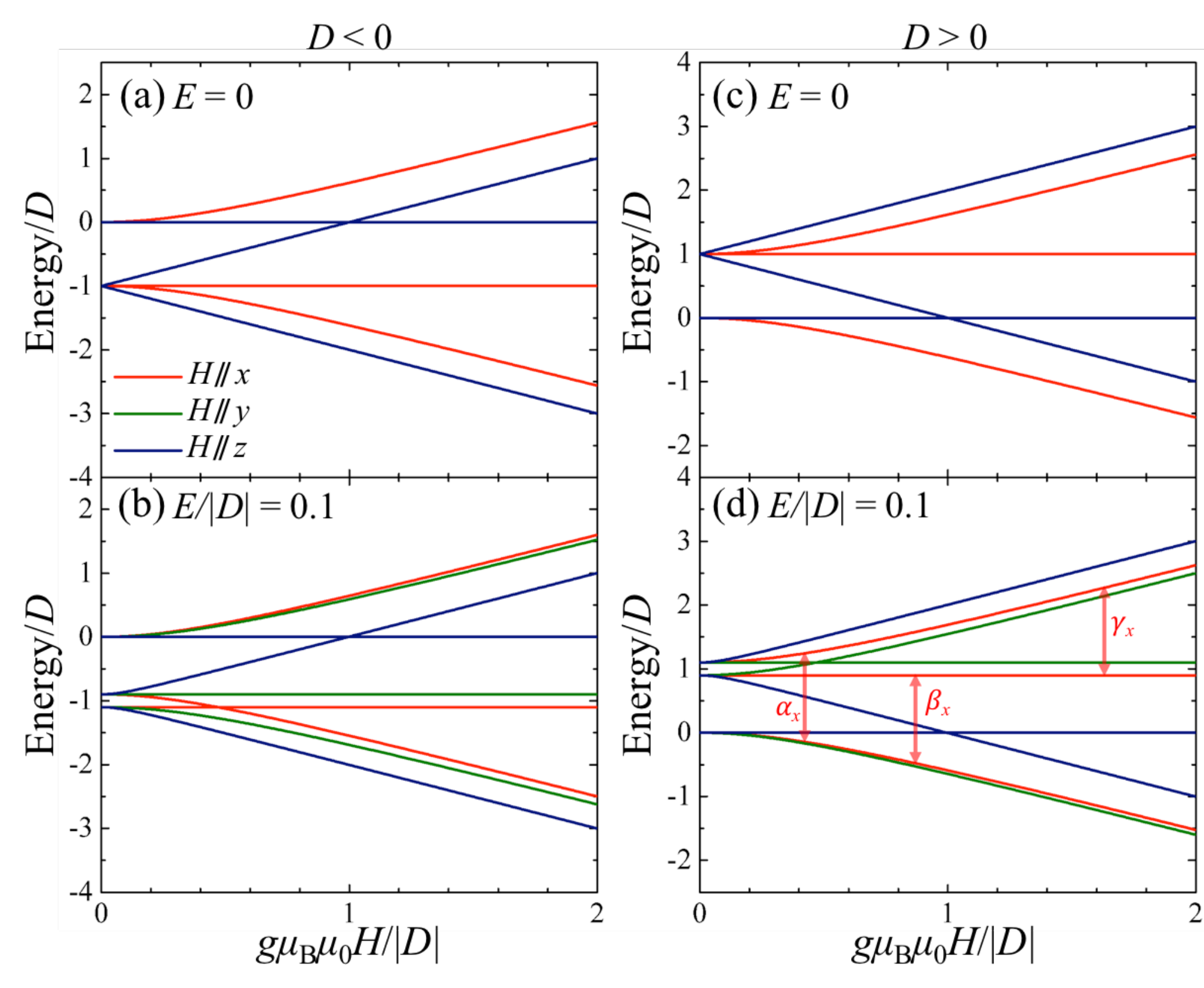}
\caption{Energy eigenvalues normalized by $D$ as a function of applied magnetic field $H$ for the Hamiltonian of Equation~\ref{ham} with isotropic $g$. In the absence of $E$ anisotropy, the levels with $H\parallel x$ and $y$ are degenerate. (a) Easy-axis scenario with $E=0$. There is no ground state level crossing for any field direction. (b) Easy axis with $E = 0.1|D|$, there is a ground state crossing for $H\parallel x$ only. (c) and (d) show the easy-plane scenario with $E=0$ and $0.1|D|$, respectively. In both a ground state level crossing occurs for $H \parallel z$. Three ESR transitions for $H\parallel x$ are labelled and discussed later.}
\label{levels}
\end{figure}

In the absence of exchange interactions, the Hamiltonian that governs
the magnetic properties in applied magnetic field is
\begin{equation}
\hat{\mathcal{H}} = D\sum_{i}(\hat{S}^z_i)^2 + E\sum_{i}[(\hat{S}^x_i)^2  - (\hat{S}^y_i)^2]
+ \mu _{\textsc{b}}\sum_i{\bf B}\cdot{\bf g}\cdot\hat{\bf S}_i, 
\label{ham}
\end{equation}
where we apply the constraint (discussed below)
\begin{equation}
0<3E<|D|. 
\label{constraint}
\end{equation}
Here $z$ is defined by the local axial direction, ${\bf g}$ is the $g$-tensor $= {\rm diag}\,(g_x, g_y, g_z)$ and $\hat{\bf S}=(\hat{S}^x, \hat{S}^y, \hat{S}^z)$ are the $S = 1$ spin operators. A negative $D$ corresponds to easy-axis anisotropy and positive $D$ is easy-plane anisotropy. 
The Hamiltonian is readily solved and the energy eigenvalues for two values of $E$ are displayed in Fig.~\ref{levels} in both the easy-axis and easy-plane cases (assuming $g$ to be isotropic). In the absence of $E$ anisotropy, the easy-axis system [Fig.~\ref{levels}(a)] has a doubly degenerate ground state that splits with applied field, and has no ground-state level crossing for any field direction. The degeneracy of $x$ and $y$ energy levels is lifted in the presence of a non-zero $E$  [Fig.~\ref{levels}(b)], and a ground-state energy level crossing appears for the field applied parallel to $x$. In contrast, for easy-plane anisotropy [Fig.~\ref{levels}(c) and (d)] a crossover from a non-magnetic to magnetic ground state occurs even if $E=0$, but only for the magnetic field parallel to $z$. Features arising from these crossovers will be observable in the polycrystalline magnetization data.

Values of $E$ outside the constraint (Eq.~\ref{constraint}) can be encompassed by an exchange of coordinate axes, without changing the system properties~\cite{boca}. For example, it can be shown (see appendix A) that a permutation of the coordinate axes leads to equivalent Hamiltonians, whose eigenvalues differ only by a constant energy shift, but which will have different values of $D$ and $E$. This means that for experiments on polycrystalline samples, where the correlation between the $z$-axis of Eq.~\ref{ham} and the crystallographic directions is lost, fitting of powder data (such as magnetic susceptibility or heat capacity, as described below) without constraining $0<3E<|D|$ can yield two apparently conflicting sets of anisotropy parameters, one with a negative axial parameter and one positive. However, only one of these sets will fulfil the constraint. The parameters can be interconverted via the relations given in appendix A.

\subsection{Effect of powder averaging on magnetometry and heat capacity measurements}

In a polycrystalline measurement of an anisotropic magnetic material the mixing of different crystal directions with respect to the applied field leads to a blurring or loss of information as compared to single-crystal experiments. However, at sufficiently low temperatures and high magnetic fields, features visible in the results of thermodynamic measurements can still yield quantitative data. Here we discuss the problem of measuring polycrystals and by looking at the results of powder-averaged simulations of isolated $S=1$ systems we show how to draw conclusions about the magnetic parameters. 
  
\subsubsection{Magnetic susceptibility}~
\newline
\noindent
It has been suggested previously that it is not possible to distinguish between easy-axis and easy-plane isolated $S=1$ magnets using polycrystalline measurements of magnetic susceptibility alone~\cite{kahn}. Revisiting this subject, we find that it is indeed possible to distinguish these two cases at sufficiently low temperatures and also extract reasonable estimates of both $D$ and $E$ from fitting polycrystalline data. It is true, however, that at high-temperatures all anisotropy information is lost.

Magnetometry measurements performed in the high-temperature, paramagnetic limit show a linear dependence of the inverse susceptibility on temperature. Extrapolating this linear dependence to obtain a temperature-axis intercept (the Weiss temperature) in isotropic magnetically-interacting systems can be used to obtain an estimate of the size of the exchange energy via the familiar Curie-Weiss law. In exchange-free, anisotropic systems the direction-dependent Weiss temperature reveals information regarding the crystal-field parameters~\cite{cho} and in particular, estimates for $D$ and $E$ could be deduced from single-crystal measurements. For example, if $E = 0$ and $g$ is isotropic, then solving the Hamiltonian above in the high-temperature region yields the Weiss temperatures $\Theta_{xy}\approx D/6$ and $\Theta_z\approx -D/3$ for the field applied perpendicular and parallel to the axial direction, respectively (see appendix B). However, in a polycrystalline experiment these values will be averaged such that the measured Weiss temperature will approach zero, and it is necessary to make measurements at lower temperatures to characterize the anisotropy.

\begin{figure}[t]
\centering
\includegraphics[width=12cm]{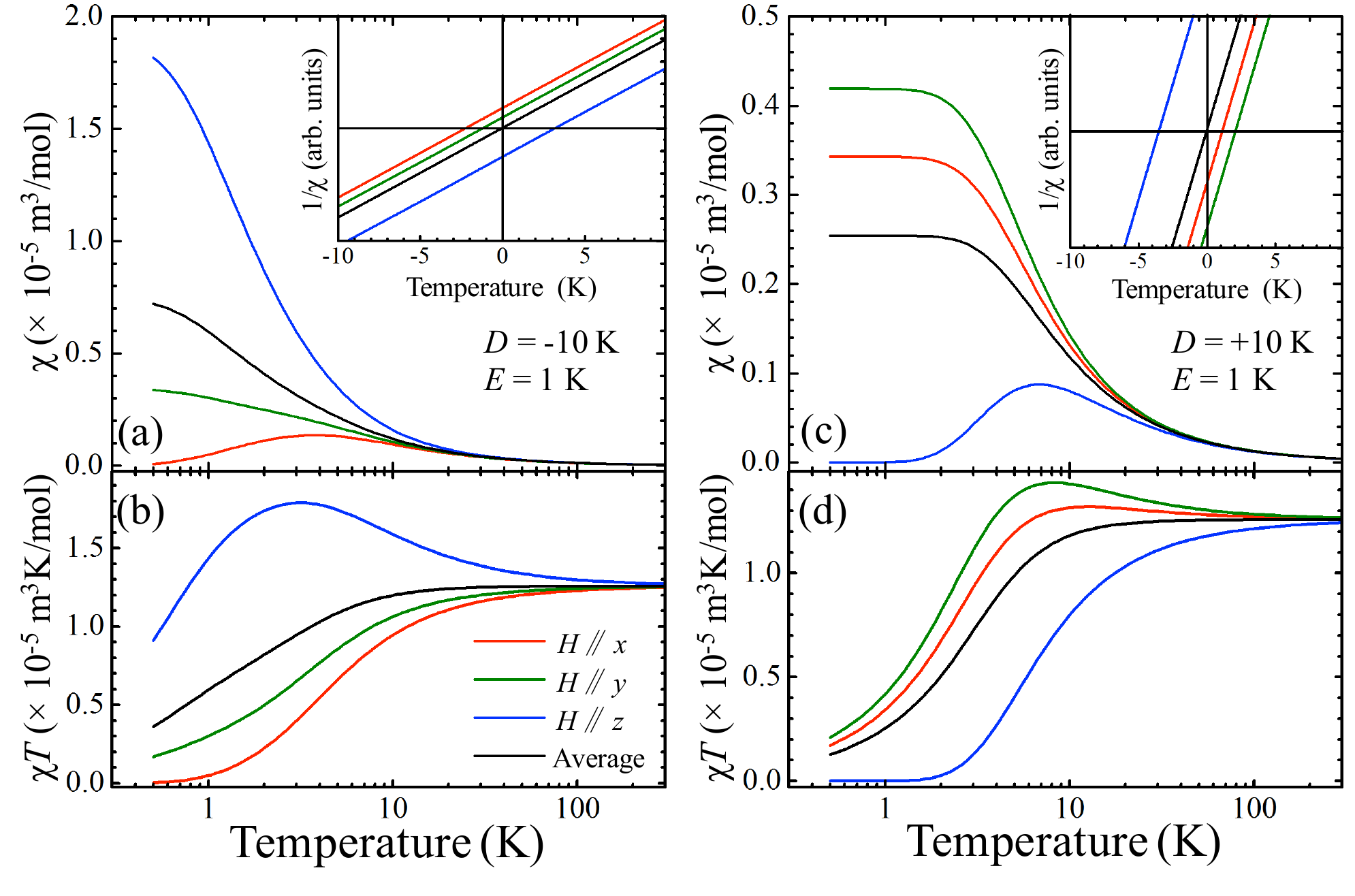}
\caption{Temperature dependence of magnetic susceptibility for different field directions calculated using Equations~\ref{sus} for $|D| = 10$~K, $E = 1$~K and isotropic $g = 2$. The simple polycrystalline average $\chi_{\rm av} = \frac{1}{3}(\chi_x +\chi_y + \chi_z)$ is also displayed. (a) and (b) show $\chi(T)$ and $\chi T(T)$, respectively, for easy-axis anisotropy. (c) and (d) show the same quantities for easy-plane anisotropy. The insets show the extrapolated values of the high-temperature inverse susceptibility, indicating the zero value of the Weiss temperature for the averaged data.} \label{sus_fig}
\end{figure}

The eigenvalues of Eq.~\ref{ham} are used to construct a partition function, from which can be found the form of the low-field molar susceptibilities for fields applied along the three principal axes~\cite{kagome}:
\begin{equation}
\begin{aligned}
\chi_x &= \frac{2N_{\rm A}\mu_0g_x^2\mu_{\rm B}^2}{D+E}~\frac{1- {\rm e}^{-\beta(D+E)}}{1+2{\rm e}^{-\beta D}\cosh{\beta E}}\\  
\chi_y &= \frac{2N_{\rm A}\mu_0g_y^2\mu_{\rm B}^2}{D-E}~\frac{1- {\rm e}^{-\beta(D-E)}}{1+2{\rm e}^{-\beta D}\cosh{\beta E}} \\ 
\chi_z &= \frac{2N_{\rm A}\mu_0g_z^2\mu_{\rm B}^2}{E}~\frac{{\rm e}^{-\beta D}\sinh{\beta E}}{1+2{\rm e}^{-\beta D}\cosh{\beta E}}
\label{sus}
\end{aligned}
\end{equation}
where $\beta = 1/k_{\rm B}T$. The expressions are in agreement with
those previously published for the case with $E\to 0$~\cite{kahn}. For these smoothly varying functions it is possible to obtain a reasonable approximation to the results of a
polycrystalline measurement from a simple average, $\chi_{\rm av} =
\frac{1}{3}(\chi_x +\chi_y + \chi_z)$. (This is in contrast to the magnetization simulations presented below, which display step-like features for certain field directions and so require averaging over more angles to reproduce the experimental data.) The success of the simple average used here will be inspected in more detail later in comparison with experimental data. 
The functions above are plotted
together with $\chi_{\rm av}$ in Fig.~\ref{sus_fig}(a) and \ref{sus_fig}(c) for the
easy-axis and easy-plane cases, respectively, with $|D| = 10$~K and $E
= 1$~K. For these values there is a clear distinction between the
easy-axis and easy-plane data, with the susceptibilities in the easy-plane case reaching a saturated value as temperature is reduced, in
contrast to the easy-axis case for which the susceptibility continues
to rise down to much lower temperatures. The insets to these figures
show the values of the inverse susceptibilities extrapolated to where
they cross the $T$-axis, highlighting the polycrystalline averaging to
zero of the Weiss temperatures. The same effect can also be seen in a
plot of $\chi T$ vs $T$ [Fig.~\ref{sus_fig}(b) and \ref{sus_fig}(d)] in which,
while the single-crystal data either strongly increase or decrease on
cooling depending on the direction of magnetic field and the sign of
$D$, the polycrystalline value remains roughly constant down to
temperatures of the order of the largest term in the Hamiltonian. In
principle, it would be possible to obtain an indication of the
size of $D$ from the temperature at which $\chi_{\rm av}T$ departs
from its high-temperature value. However, more reliable estimates can
be obtained by direct fitting of experimental data to $\chi_{\rm av}$
as described below.

\begin{figure}[t]
\centering
\includegraphics[width=12cm]{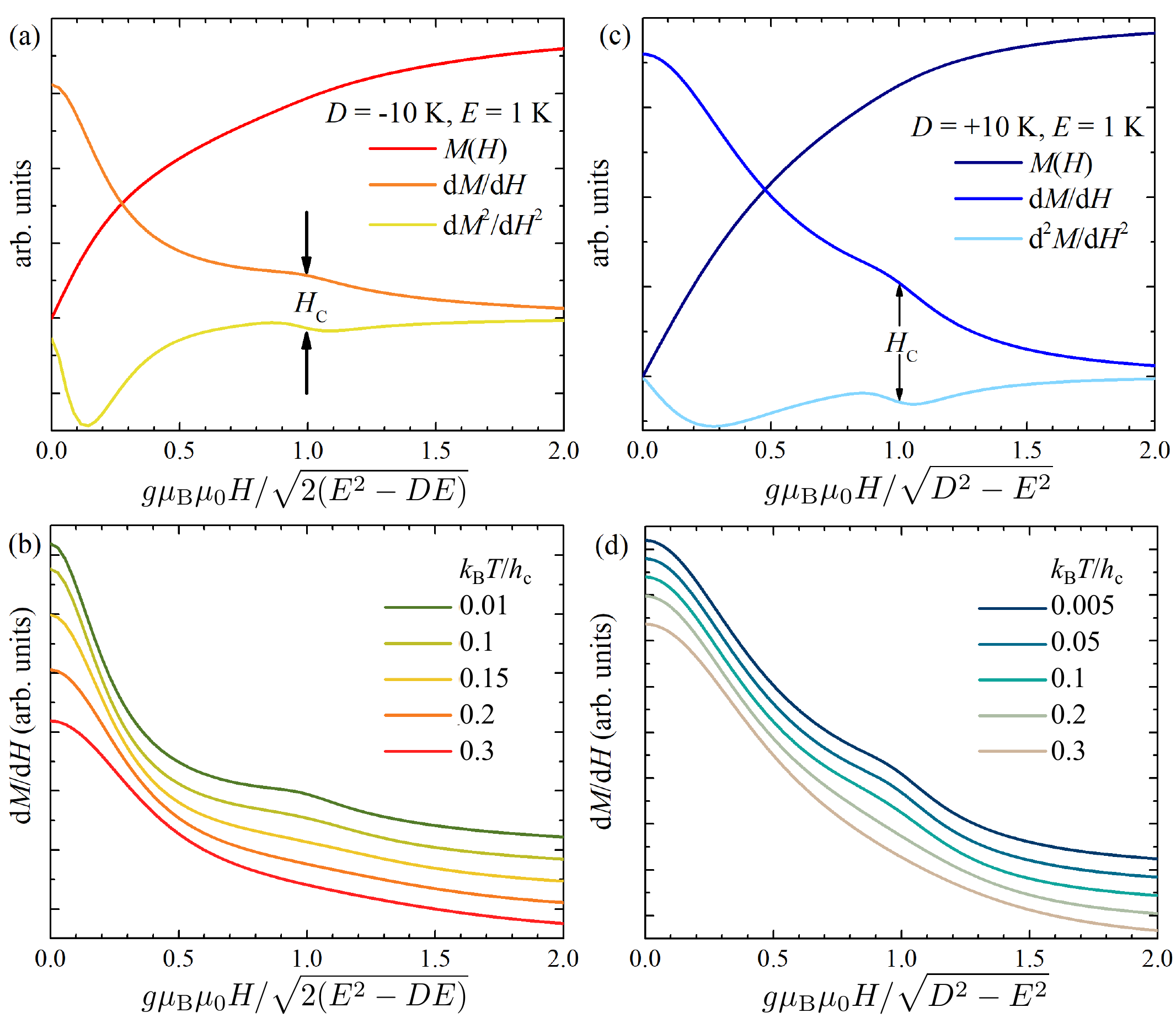}
\caption{Simulated full polycrystalline magnetization $M$ and associated gradients for $|D| = 10$~K $E = 1$~K and isotropic $g = 2$. (a) and (b) show the simulations for easy-axis anisotropy revealing a bump in ${\rm d}M/{\rm d}H$ at $H_{\rm c}$ given by Equation~\ref{hc_ezaxis}, due to a ground state level crossing for fields parallel to $x$. (c) and (d) show the simulations for easy-plane anisotropy. A feature is observed at a critical field defined by Equation~\ref{hc_ezplane}. In both cases the feature can only be observed at thermal energies low compared to $h_{\rm c} = g\mu_{\rm B}\mu_0H_{\rm c}$. The data in (b) and (d) are offset for clarity.}  \label{mag_fig}
\end{figure}

\subsubsection{Magnetization}~
\newline
\noindent
At sufficiently low temperatures the magnetization $M_i$ for fields applied parallel to $i = x$, $y$ and $z$ will be dominated by the ground state energy level crossings seen in Fig.~\ref{levels}. As such, for easy-axis anisotropy a step will be observed in $M_x$ given by
\begin{equation}
g\mu_{\rm B}\mu_{0}H_{\rm c} =\sqrt{2(E^2-DE)},
\label{hc_ezaxis}
\end{equation}
which is zero for $E = 0$, while for easy-plane anisotropy there will be a step in $M_z$ at a critical field given by 
\begin{equation}
g\mu_{\rm B}\mu_{0}H_{\rm c} =\sqrt{D^2-E^2}.
\label{hc_ezplane}
\end{equation}
The abrupt changes in $M$ at the critical fields mean that the simple average used above for susceptibility will not reliably reproduce the result of a measurement. Instead we perform a simulation of polycrystalline $M(H)$ using a full angular average over many possible field directions~\cite{SI}, the results of which are shown in Fig.~\ref{mag_fig}(a) and (c) for easy-axis and easy-plane cases respectively. The easy-axis (easy-plane) system will have a sharp increase in $M_x$ ($M_z$) at $H = H_{\rm c}$ and hence a peak in the differential susceptibility. For a powder, this feature is reduced to a small bump, which is hard to discern in $M$, but is readily observed in ${\rm d}M/{\rm d}H$ or ${\rm d^2}M/{\rm d}H^2$ as seen in the figures. Thermal occupation of excited states obscures the crossing of the ground state and the strength of the features diminishes as temperature is raised. By simulating the differential susceptibility at different temperatures [Fig. \ref{mag_fig}(b) and (d)], we find that the peak indicating the level crossing at $H_{c}$ can be observed only if the temperature is lowered below approximately $0.1\times g\mu_{\rm B}\mu_0H_{\rm c}/k_{\rm B}$ in both cases.

\begin{figure}
\centering
  \includegraphics[width=14cm]{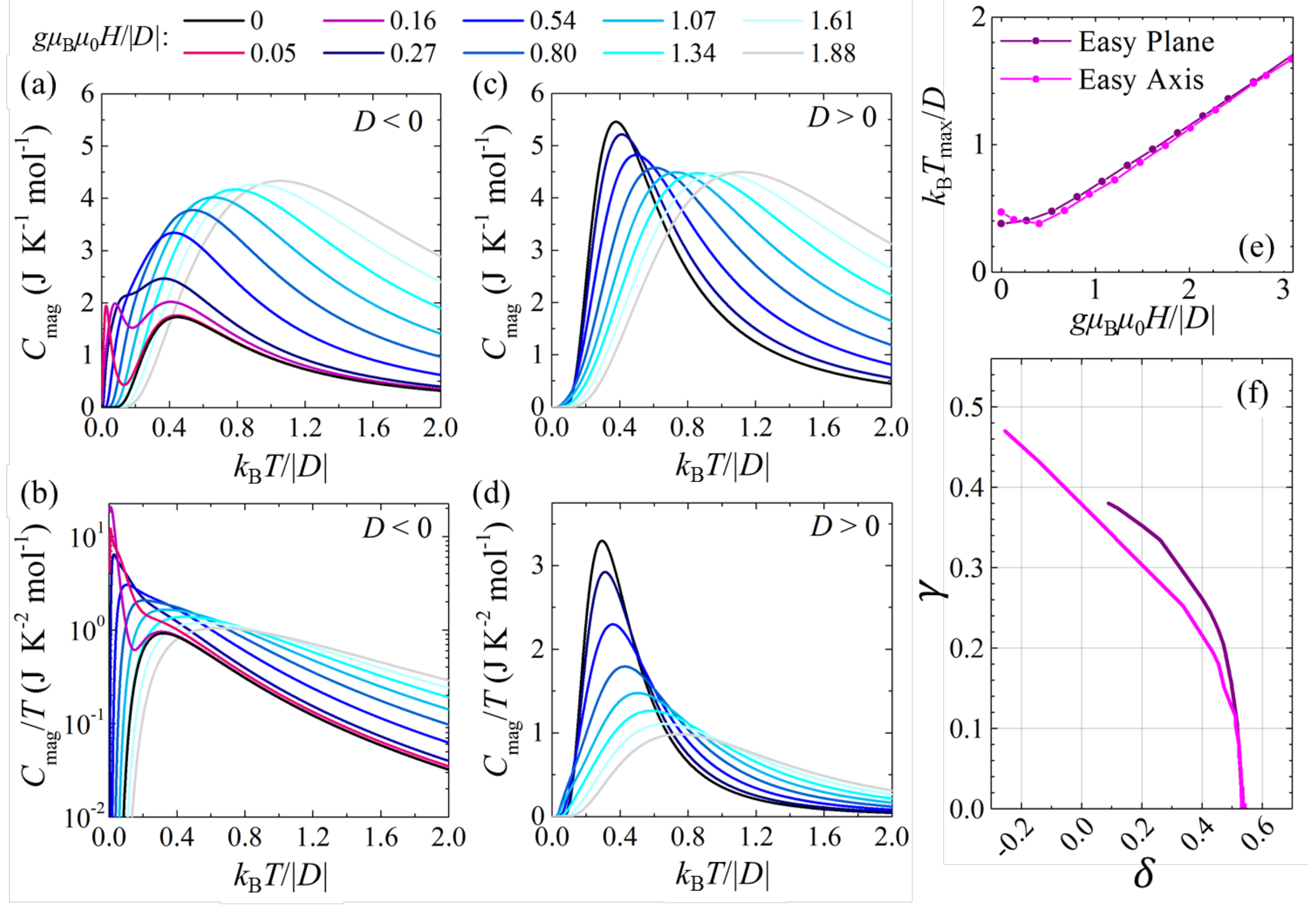}
  \caption{(a) and (b) show simulations with $E=0$ of magnetic heat capacity $C_{\rm{mag}}$ and $C_{\rm{mag}}/T$ as a function of temperature at various fields for easy-axis anisotropy. (c) and (d) show the same simulated quantities for easy-plane anisotropy. (e) The position $T_{\rm{max}}$ of the broad maximum in the simulated $C_{\rm mag}$, plotted against field in dimensionless units. (f) Correlation between the parameters of Equation~\ref{grad} resulting from local linear fitting of the $T_{\rm{max}}$ vs $H$ data.}
  \label{hcap_sim}
\end{figure}

\subsubsection{Heat capacity}~
\label{hc_theory}
\newline
\noindent
The zero-field magnetic heat capacity $C_{\rm mag}$ (in units of $N_{\rm A}k_{\rm B}$) resulting from solving Equation~\ref{ham} is found to be
\begin{equation}
C_{\rm mag}(T) = \frac{2{\rm e}^{\beta D}(D^2+E^2)\cosh{\beta D}+4E(E - D{\rm e}^{\beta D}\sinh{\beta D})}{(k_{\rm B}T)^2({\rm e}^{\beta D}+2\cosh{\beta E})^2},
\label{zf_hc}
\end{equation}
and is in agreement with the expression previously published for the case with $E\to 0$~\cite{boca}. As temperature is reduced the function reproduces anomalies in the heat capacity resulting from the zero-field splittings shown in Fig.~\ref{levels}, and can be used in combination with lattice heat capacity models to fit measured single or polycrystalline data as shown by example later. 

The evolution of $C_{\rm mag}(T)$ under applied field for a polycrystalline sample is distinct for the easy-axis and easy-plane cases. Simulations of $C_{\rm mag}(T)$ for both situations are obtained via a full polycrystalline average at various fields with $E = 0$~\cite{SI} and displayed in Fig.~\ref{hcap_sim}. For the easy-axis scenario $C_{\rm mag}(T)$  [Fig.~\ref{hcap_sim}~(a)] shows a single broad maximum at a temperature set by $D$. In an applied field, the ground state degeneracy is lifted, resulting in the emergence of a second narrow peak at low temperatures. As the field is raised the two peaks merge and move to higher temperatures while increasing in amplitude and breadth. For easy-plane anisotropy [Fig.~\ref{hcap_sim}~(c)] only a single broad peak is apparent, which initially drops in amplitude, gets broader and moves to higher temperatures as the field is applied. The shift in magnetic entropy from low to high temperatures caused by the field-induced splitting of energy levels can been appreciated from the form of the $C_{\rm mag}/T$ curves shown in Fig.~\ref{hcap_sim}(b) and \ref{hcap_sim}(d). 

Another estimate of the size of $D$ in systems with $E=0$ can be obtained from the field dependence of the position, $T_{\rm{max}}$, of the broad maximum observed in $C_{\rm mag}(T)$. The values obtained from the simulated data are plotted in dimensionless units in Fig.~\ref{hcap_sim}(e), and can be parametrised as follows,
\begin{equation}
\frac{k_{\rm{B}}T_{\rm{max}}}{|D|} = \gamma + \delta\left(\frac{g\mu_{0}\mu_{\rm{B}}H}{|D|}\right),
\label{grad}
\end{equation}
with two field-dependent parameters, $\delta$---the local gradient and $\gamma$---the local, extrapolated zero-field intercept. The correspondence between $\delta$ and $\gamma$ is shown in Fig.~\ref{hcap_sim}(f). As will be shown later for experimental data, an estimate of $\gamma D$ and $\delta$ can be found from a linear fit to the measured values of $T_{\rm{max}}$ vs $g\mu_{0}\mu_{\rm B}H$, while the pre-factor $\gamma$ can be uniquely determined for a particular sign of $D$ by using the fitted value of $\delta$ and Fig.~\ref{hcap_sim}(f). 

\subsection{Experimental results for \Nilut}

\subsubsection{Crystal structure}
\begin{figure}[t]
\centering
\includegraphics[width=11.0cm]{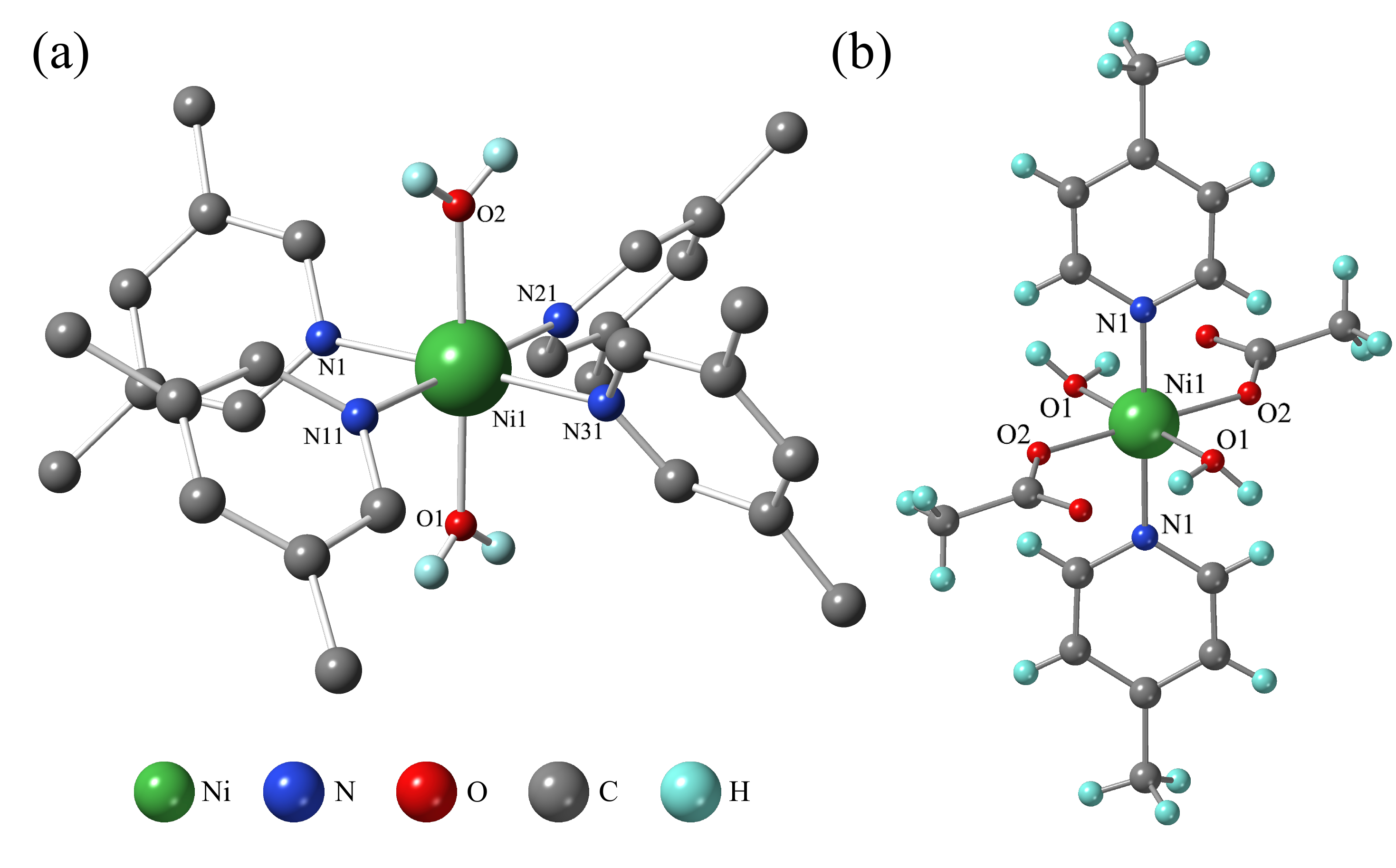}
\caption{(a) Local Ni(II) environments determined by single-crystal x-ray diffraction of (a) \Nilut~at $T = 100$~K, and (b) \Niace~at $T = 150$~K. In (a) lutidine hydrogens and BF$_{4}^{-}$ counter ions are omitted. The stacking of the molecular units in both materials is shown in~\cite{SI}.} 
\label{mol_struc}
\end{figure}

~
\newline
\noindent
\Nilut~crystallizes in the monoclinic space group $P2_{1}/n$. Fig.~\ref{mol_struc}(a) shows the coordination environment deduced from single-crystal synchrotron x-ray diffraction performed at 100~K, and structural parameters are found in Table~\ref{tab_struc}. The crystallite used was of the order of $50\times50\times50$~$\mu$m$^3$; sufficient for the structural studies, but too small for thermodynamic measurements. The material is made up of distorted NiN$_{4}$O$_{2}$ octahedra with four equatorial nitrogens donated by 3,5-lutidine, and two axial oxygens provided by water. The three bond angles between opposite donor atoms in the nickel octahedra are within the range $176.95^{\circ}$--$179.17^{\circ}$, and the {\it cis} N---Ni---O angle ranges between $87.7^{\circ}$ and $92.3^{\circ}$. 

The Ni(II) ions in the [Ni(H$_{2}$O)$_{2}$(3,5-lut)$_{4}$] complexes are well-isolated by the non-bridging lutidine molecules and adjacent complexes are kept apart by two BF$_{4}^{-}$ counter ions, which are hydrogen bonded to the water molecules. The nearest Ni---Ni neighbours are separated by approximately 9.2~\AA\ along the [101] crystal direction. As a result, the exchange interactions between the $S = 1$ Ni(II) ions are expected to be negligible and the low-temperature magnetic properties should be dominated by single-ion anisotropy. The breaking of four-fold rotational symmetry by the equatorial ligands suggests that $E$ will be non-zero.

 \begin{table}[t]
\centering
\begin{tabular}{|c|l|l|l|}
\hline
Compound				& Ni(lut) 			& Ni(ace)			& Ni(pyz)  			\\ \hline							
$T$ (K) 					& 100		 		& 150		      		& 300		 		\\ 
crystal system			&  monoclinic		& orthorhombic		&  tetragonal   		\\ 
space group 			& $P2_{1}/n$  		& $Pcab$      		& $I 4/m c m$ 		\\ 
$a$ (\AA)				&12.2611(8)			& 8.8996(3)   		& 9.91670(18)		\\ 
$b$ (\AA)				&17.0125(12)  		& 12.3995(4)  		& 9.91670(18)		\\ 
$c$ (\AA)				&16.7006(11)  		& 17.6516(7)  		& 14.8681(4)		\\ 
$\beta$ ($^\circ$)			& 103.416(1)			& 90.00				& 90.00		\\
Ni---~(\AA)   			& O1 = 2.099(2) 		& N1 = 2.107(3)  	& O1 = 2.050(7) 	  	\\ 
Ni---~(\AA)   			& O2 = 2.08(2)  		& N1 = 2.107(3)  	& O2 = 2.050(7)	  	\\
Ni---~(\AA)   			& N1 = 2.110(3) 		& O1 = 2.073(2)  	& N1 =  2.1724(18)  	\\ 
Ni---~(\AA) 		 		& N11= 2.105(3) 	& O1 = 2.073(2) 		& N11= 2.1724(18)   	\\ 
Ni---~(\AA)  				& N21= 2.094(3) 	& O2 = 2.059(2) 		& N21= 2.1724(18)   	\\ 
Ni---~(\AA)  				& N31= 2.115(3) 	& O2 = 2.059(2) 		& N31= 2.1724(18)   	\\ \hline                                  
\end{tabular}	
\caption{Structural parameters and local environment at temperature $T$ of the three compounds discussed in this paper. \mbox{Ni(lut)~=~\Nilut,} \mbox{Ni(ace)~=~\Niace,} and \mbox{Ni(pyz)~=~\Nipyz.} Atomic labels are shown in Fig.~\ref{mol_struc}~and~\ref{pyz_struc}.}
\label{tab_struc}	
\end{table}

\begin{figure}[t]
\centering
\includegraphics[width=12.0cm]{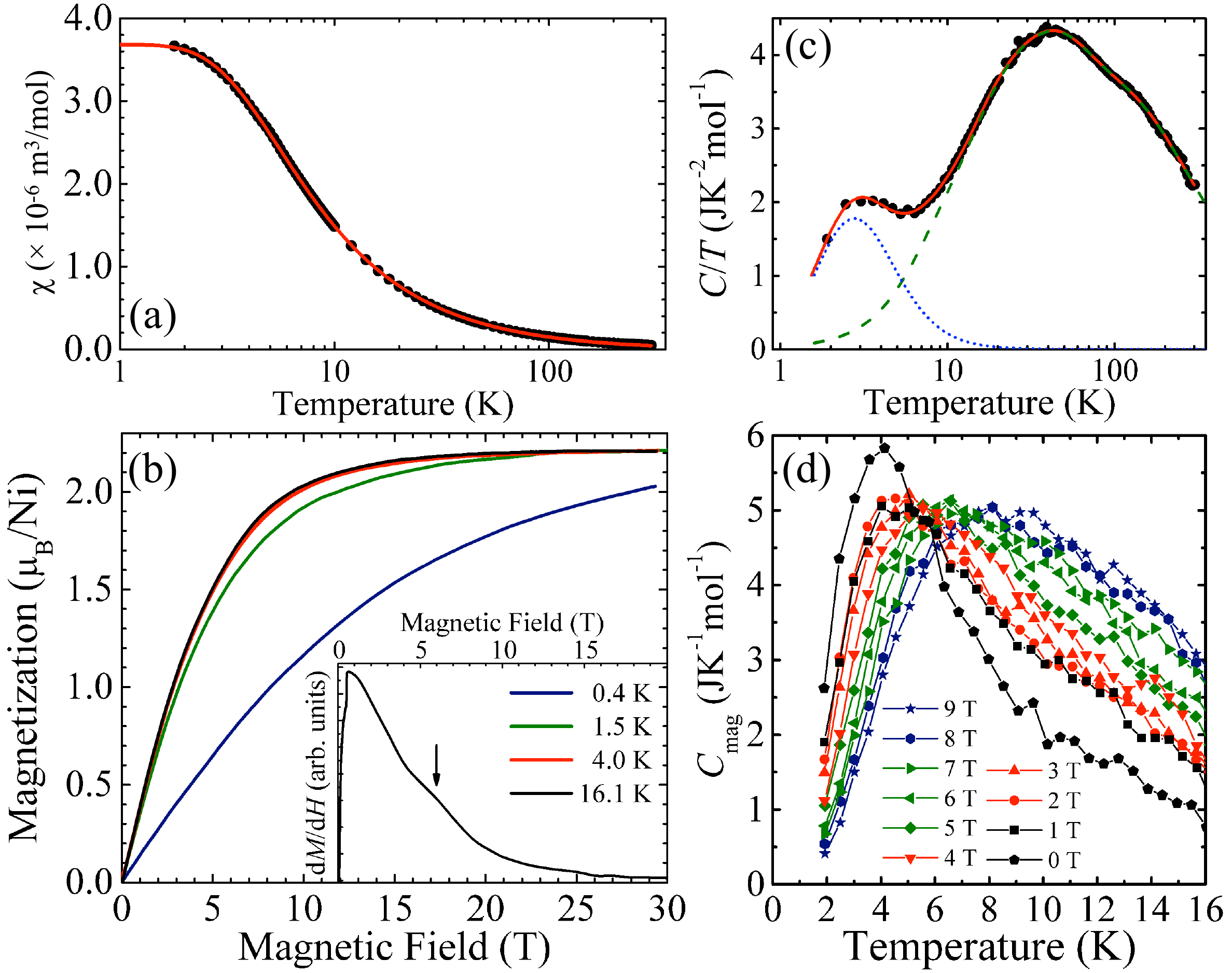}
\caption{Polycrystalline thermodynamic properties of \Nilut. (a) Magnetic susceptibility $\chi(T)$ measured at $\mu_0H=0.1$~T (circles). The line is a fit to the simple polycrystalline average described in text. (b) Magnetization measured at the temperatures shown. The inset shows the differential susceptibility ${\rm d}M/{\rm d}H$ measured at 0.4~K. The critical field $\mu_0 H_{\rm c}$ is indicated by an arrow. (c) Measured heat capacity divided by temperature (circles). The solid red line is a fit to the lattice plus magnetic model described in the text. The dashed green line is the lattice contribution and the dotted blue line is the magnetic part. (d) The magnetic heat capacity $C_{\rm mag}(T)$ at various fields, obtained by subtracting the zero-field lattice contribution from the measured data.} 
\label{lut_data}
\end{figure}

\subsubsection{Thermodynamic measurements}~
\newline
\sloppypar
\noindent
The magnetic susceptibility of a polycrystalline sample of \Nilut~
measured at $\mu_0H=0.1$~T is shown in Figure~\ref{lut_data}(a) and
resembles the data for an ensemble of $S = 1$ moments with single-ion
anisotropy, but no significant exchange interactions. The data are
fitted with the function $\chi(T) = \chi_{\rm av}(T) + \chi_0$, under
the constraint $0<3E<|D|$. Here $\chi_{\rm av}$ is the simple
polycrystalline average $\frac{1}{3}(\chi_x +\chi_y + \chi_z)$ of the
susceptibility components defined by Equation~\ref{sus}, and $\chi_0$
is a temperature-independent contribution. Successful fitting of the
data requires using the approximation that $g_x=g_y=g_z=g$ in
Equation~\ref{sus}. In reality this will not be the case; the presence
of single-ion anisotropy suggests that the components of $g$ are
unequal, as the same effects give rise to both. Perturbation theory
predicts that $g_z - g_{xy} = 2D/\lambda$ and $g_x-g_y = 4E/\lambda$,
where $\lambda$ is the spin-orbit coupling, which is $\sim -500$~K for
Ni(II) in octahedral environments~\cite{boca}. However, in the types
of system we consider here, typical values are $D\sim10$~K and $E\sim1$~K,
so that $\Delta g$ is expected to be $\sim0.01$. Thus the uniform $g$
approximation is reasonable within the errors of the thermodynamic
measurements. The parameters resulting from the fit to $\chi(T)$ are
$g = 2.24(1)$, an easy-plane $D = 8.7(2)$~K, $E = 1.2(2)$~K and $\chi_0 =
-8(1)\times10^{-9}$~m$^3$mol$^{-1}$.\footnote{Note that a fit
  performed without constraining the upper limit of $E$ can also yield
  $D = -6.3(5)$~K and $E = 3.7(1)$~K, with $g$ and $\chi_0$ similar to
  the constrained fit. These values for $D$ and $E$ do not obey
  $3E<|D|$ and are the result of the permutation of the coordinate
  axes described earlier. They reduce to the values obtained from the
  constrained fit under the transformations of Eq.~\ref{convert}.}

\vspace{0.3cm}
$M(H)$ data measured at various temperatures are shown in Fig.~\ref{lut_data}(b). The data increase smoothly towards to the saturated value of $2.21(2)\mu_{\rm B}$ per Ni(II), which is consistent with the polycrystalline averaged value of the $g$ factor resulting from fitting $\chi(T)$. The lowest temperature curves show a kink in $M(H)$ close to 5~T. This is more clear on differentiating the 0.4~K data (inset), where it appears as a small bump resembling the feature discussed earlier that arises from a ground state energy level crossing. The position of the bump is $\mu_0 H_{\rm c} = 6.0(6)$~T. Using the easy-plane model (Eq.~\ref{hc_ezplane}) an estimate of $\sqrt{D^2-E^2} = 9.0(9)$~K is obtained, which is in agreement with the susceptibility results.\footnote{We note that, while these $M(H)$ data are collected using pulsed magnetic fields, the location of $H_{\rm c}$ in this case is within the field and temperature range of more conventional magnetometers equipped with a $^3$He refrigerator.}

\vspace{0.3cm}
Zero-field heat capacity measurements of polycrystalline \Nilut\ are shown as $C/T$ vs~$T$ in Fig.~\ref{lut_data}(c). The data exhibit two peaks, one around 40~K due to the phonons, and a second at $\approx 3$~K which is attributed to single-ion anisotropy. The proximity of lattice and magnetic contributions to the heat capacity mean that dealing with each separately is not possible. Instead we fit the data to a model $C/T = C_{\rm latt}/T + C_{\rm mag}/T$, where $C_{\rm latt}$ approximates the lattice contribution using a model with one Debye and three Einstein phonon modes~\cite{SI,mansonjacs2009}, and $C_{\rm mag}$ is given in Eq.~\ref{zf_hc}. The fit is shown in the figure as a solid red line and is seen to account well for the data across the whole temperature range. Also shown are the separate lattice (dashed line) and magnetic contributions to the fit. The anisotropy parameters resulting from the fit are easy-plane $D = 10.4(1)$~K and $E = 2.6(2)$~K.\footnote{The fit of the lattice contribution yields the following characteristic amplitudes, $A_i$~(J\,K$^{-1}$\,mol$^{-1}$), and temperatures, $\theta_i$~(K), of the Debye ($i=$~D) and Einstein ($i=$~E) phonon modes: $A_{\rm D} = 53(3)$, $\theta_{\rm D} = 50(1)$, $A_{\rm E1} = 128(5)$, $\theta_{\rm E1} = 87(3)$, $A_{\rm E2} = 199(5)$, $\theta_{\rm E2} = 195(7)$, $A_{\rm E3} = 388(6)$ and $\theta_{\rm E3} = 540(9)$.} These values are in agreement with $\sqrt{D^2-E^2} = 9.0(9)$~K estimated from the magnetization data. The size of $D$ is similar to that obtained from the fits of the susceptibility to the simple polycrystalline average model, while $E$ differs by 50\% from the susceptibility value. 

Heat capacity measurements were also performed at fixed values of applied field $0\le\mu_0 H\le9$~T. The lattice contribution determined from the fit to the zero-field data is subtracted from each trace to yield $C_{\rm mag}(T)$ at different fields, which are shown in Fig.~\ref{lut_data}(d). It is seen that the broad hump due to energy level splitting initially drops in amplitude as the field is turned on, and at higher fields broadens and shifts to higher temperatures. This is very similar to the behaviour of the simulated data shown in Fig.~\ref{hcap_sim}(c), further confirming the easy-plane nature of this material.  

\begin{figure}[t]
\centering
\includegraphics[width=12.0cm]{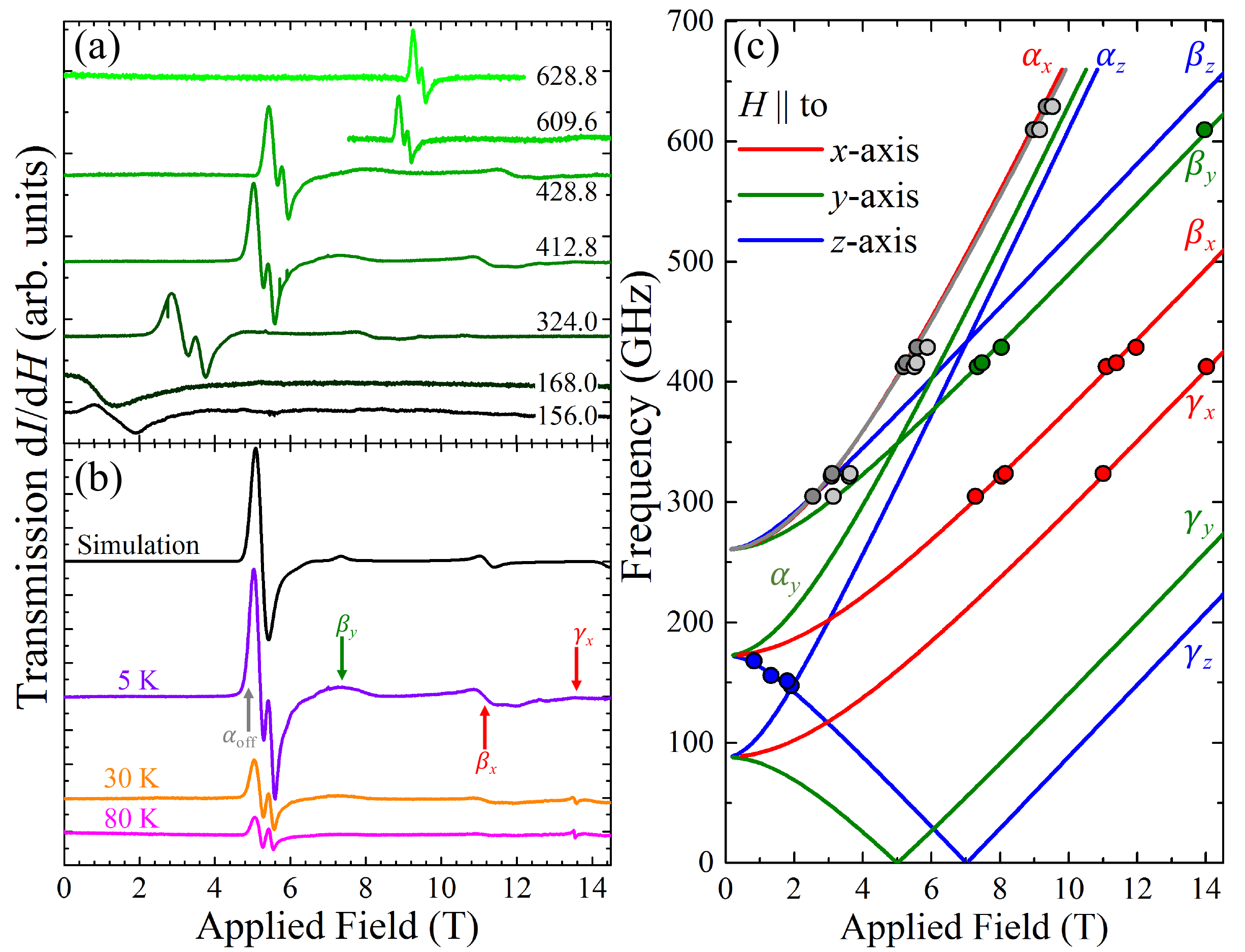}
\caption{High-frequency ESR results for polycrystalline \Nilut. (a) Representative spectra in first derivative mode collected at 5~K with various frequencies. (b) Temperature dependence of the 412.8~GHz spectra. The transitions are labelled according to the description in the text. The black line is a simulation at 5~K using the parameters derived from the fit in the next panel. (c) Frequency versus field plot showing peak positions (circles) observed at 5~K. The lines are expected locations of the resonances deduced from a fit to the experimental data described in the text, with the grey lines and circles arising from off-axis resonances.} 
\label{lut_esr}
\end{figure}

\subsubsection{Electron-spin resonance}~
\newline
\noindent
In order to check the validity of our proposed methodology for determining anisotropy parameters from thermodynamic measurements on polycrystalline $S=1$ systems, we also investigated \Nilut~using high-frequency ESR. Arguably, ESR is the technique best suited to the evaluation of single-ion anisotropy in a powdered sample. However, for all but the smallest zero-field splittings, the frequency and field regimes needed to observe the crucial transitions requires highly specialised, non-standard equipment. Our measurements were performed at the National High Magnetic Field Laboratory, Tallahassee, Florida~\cite{SI}. 

In a polycrystalline $S = 1$ sample with single-ion anisotropy, multiple ESR transitions are expected between the split triplet energy levels. At a given frequency, for the field applied parallel to the local $i$-axis ($i = x$, $y$ or $z$), there are two transitions possible that obey the ESR selection rule, $\Delta m_s = \pm 1$: one at low field and one at high field, which we label $\beta_i$ and $\gamma_i$, respectively. In addition, it is also possible to see an excitation with $\Delta m_s = \pm 2$, this so-called {\it half-field transition} is labelled $\alpha_i$. Formally such transitions are forbidden, but when the Zeeman energy is comparable to the zero-field splitting strong mixing between $m_s$ states occurs and the selection rule is relaxed. Examples of the $\alpha_x$, $\beta_x$ and $\gamma_x$ ESR transitions are indicated in Fig.~\ref{levels}(d). Additional lines may also be observed at positions that do not correspond to one of the Cartesian axes. These {\it off-axis} resonances may be present at the half-field transitions and have been known to dominate the polycrystalline spectra~\cite{off1,off2,off3}.    

ESR spectra were recorded in first derivative mode at frequencies in the range $100<\nu<630$~GHz at 5~K and representative data are shown in Figure~\ref{lut_esr}(a). A broad feature is observed around 1~T in the 156.0 GHz data that drops in field as the frequency is raised and is attributed to the $\gamma_z$ transition. At higher frequencies a large double resonance is observed (e.g. near 5~T at 412.8~GHz), which sharpens and moves to high fields with increasing frequency. The larger of these two peaks is attributed to the off-axis half-field transition $\alpha_{\rm off}$, and is expected to lie very close to the $\alpha_x$ transition. The broad hump at slightly higher fields is ascribed to $\beta_y$ transitions. At yet higher fields smaller features are seen that can be attributed to the $\beta_x$ and $\gamma_x$ transitions. 

Transitions are labelled in the temperature-dependent spectra recorded at 412.8~GHz and shown in Fig.~\ref{lut_esr}(b). Both the $\alpha_{\rm off}$ line and the $\beta_x$ line increase in amplitude as temperature is reduced, identifying them as excitations from the ground state. The $\gamma_x$ transition is seen to be smaller than $\beta_x$ at low temperatures and diminishes in amplitude when the temperature falls, which is expected for a transition between excited states. These observations identify the energy-level splitting as easy plane.

The frequency and field positions of the transitions are modelled with an easy-plane energy-level scheme in Figure~\ref{lut_esr}(c). Fitting is performed as described in Ref.~\cite{esrfit} and the best fit to the data is found for the parameters $g_x = 2.21(1)$, $g_y = 2.17(1)$, $g_z = 2.16(3)$, $D = 10.40(1)$~K and $E = 2.11(4)$~K. The fit successfully reproduces most of the peak positions and, as shown in panel (b), a simulation of the 412.8~GHz spectrum arising from these parameters compares reasonably well with the measured data. Note that $g_x > g_z$, which is consistent with the results of perturbation theory for a Ni(II) ion with easy-plane anisotropy~\cite{boca}. The observed splitting of the off-axis half-field transition is not explained by the simulations. One possible reason for the extra peak could be the presence of a second Ni(II) site in the ESR sample with slightly different anisotropy parameters,\footnote{$g_x = g_y = 2.20$, $g_z = 2.16$, $D = 9.2$~K and $E = 2.08$~K.} however no evidence of a significant impurity fraction is observed in the synchrotron x-ray diffraction measurements. Another possible explanation is the presence of small spin-spin couplings between the Ni(II) ions. The crystal structure does not show any evidence for significant exchange pathways and any magnetic interactions must be less than $\sim 1$~K or their effect would be observed in the magnetometry measurements, but very small couplings have previously be found to lead to ESR peak splitting in molecular systems, even at elevated temperatures~\cite{hill}. 

Whatever the explanation for the extra peak, the best fit $D$ and $E$ parameters account for the majority of the ESR resonances and are in excellent agreement with the values derived from the heat capacity analysis and the position of $H_{\rm c}$ in the low temperature $M(H)$ data. 

\subsubsection{Discussion}~
\newline
\noindent
In light of the high-frequency ESR data we can judge the effectiveness of the analysis methodology for magnetometry and heat capacity. The thermodynamic measurements are all strongly indicative of an easy-plane anisotropy in this material, which is confirmed by ESR. The $D$ and $E$ parameters derived from fitting to the zero-field heat capacity agree closely with those obtained from ESR. The agreement is less good for the parameters deduced from fitting susceptibility data. The fitting function in this case makes use of the elementary polycrystalline average  $\chi_{\rm av} = \frac{1}{3}(\chi_x +\chi_y + \chi_z)$, which simplifies the fitting procedure, but perhaps does not sample enough field angles to fully account for the data. Nevertheless, the estimate of the size of the parameters obtained from the susceptibility matches the ESR and heat capacity values to within less than 20\% for $D$ and 50\% for $E$.

Owing to the polycrystalline nature of the sample, it is not possible to identify the easy plane of \Nilut~from the thermodynamic or ESR measurements alone. However the symmetry of the NiN$_{4}$O$_{2}$ octahedra would strongly suggest that the hard $z$-axis is perpendicular to the NiN$_{4}$ equatorial plane.

\subsection{Experimental results for \Niace}

\subsubsection{Crystal structure}~
\newline
\noindent
Having introduced the analysis methods with the previous easy-plane material, we now test them on a Ni(II) material with a different local environment. \Niace~crystallizes in the orthorhombic space group $Pcab$. Fig.~\ref{mol_struc}(a) shows the coordination environment deduced from single-crystal x-ray diffraction performed at 150~K on a microcrystal, and structural parameters are found in Table~\ref{tab_struc}. The material contains distorted NiO$_4$N$_2$ octahedra, as compared to the NiN$_4$O$_2$ octahedra in the system discussed above. In the present case, the local environment is made up of two axial nitrogen atoms donated by 4-picoline and four equatorial oxygen atoms, two donated by acetate and two from water. The three bond angles between opposite donor atoms in the nickel octahedra are all $180^{\circ}$, and the {\it cis} O---Ni---N angle ranges between $87.3^{\circ}$ and $92.7^{\circ}$.

The individual \Niace~molecular units are well-separated in the $c$-direction by the 4-picoline molecules. The closest Ni---Ni distance is approximately 7.6~\AA~within the $ab$-plane, but with no apparent exchange pathway between nearest neighbours. Hence the magnetic properties are again expected to be that of an ensemble of magnetically isolated $S = 1$ moments with single-ion anisotropy.

\subsubsection{Thermodynamic measurements}
\begin{figure}[t]
\centering
\includegraphics[width=12.0cm]{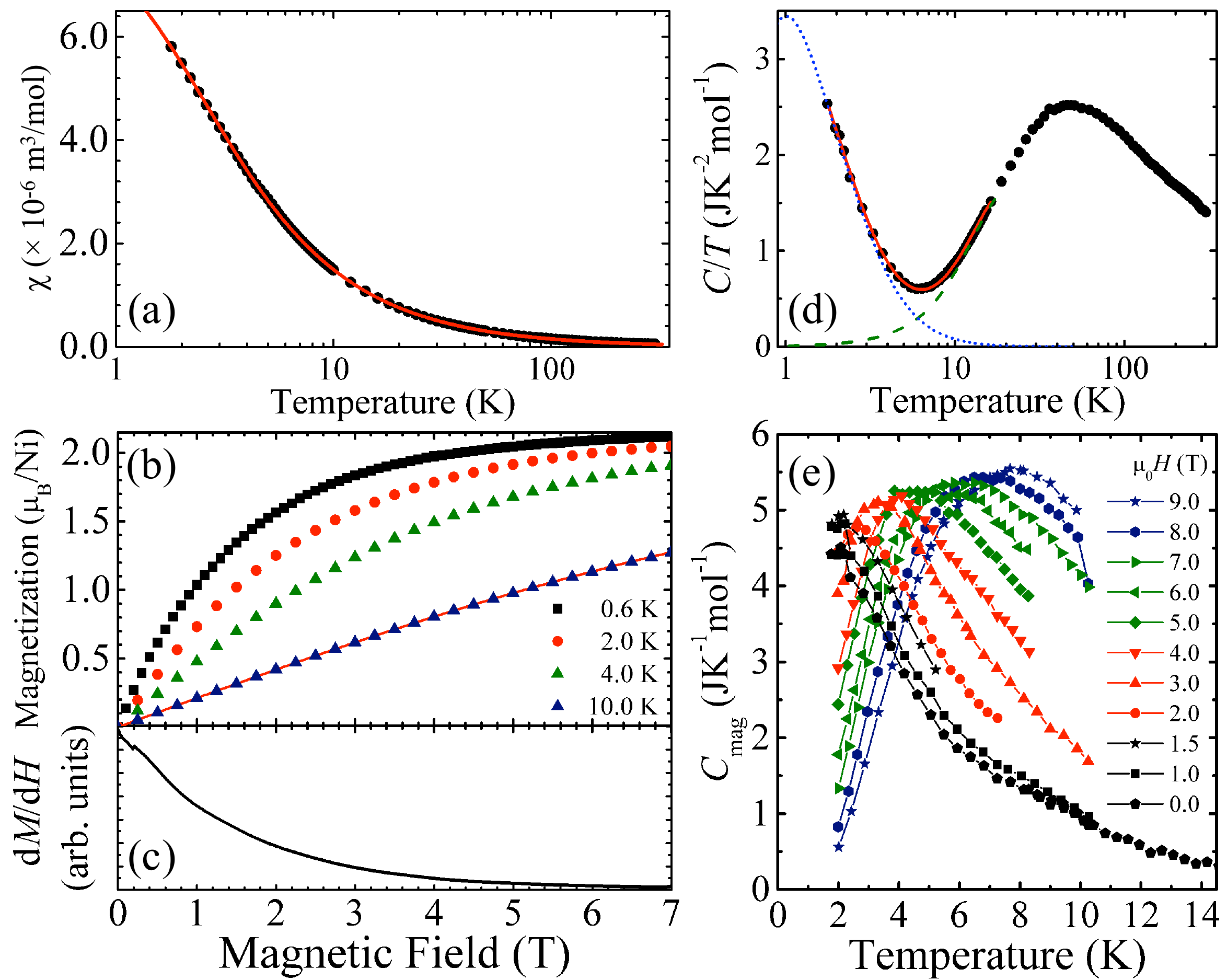}
\caption{Polycrystalline thermodynamic properties of \Niace. (a) Magnetic susceptibility $\chi(T)$ measured at $\mu_0H=0.1$~T (circles). The line is a fit to the simple polycrystalline average described in the text. (b) Magnetization measured at the temperatures shown. (c) Differential susceptibility ${\rm d}M/{\rm d}H$ measured at 0.6~K. (d) Measured heat capacity divided by temperature (circles). The solid red line is a fit to the lattice plus magnetic model described in the text. The dashed green line is the lattice contribution and the dotted blue line is the magnetic part. (e) The magnetic heat capacity $C_{\rm mag}(T)$ at various fields, obtained by subtracting the zero-field lattice contribution from the measured data.} 
\label{ace_data}
\end{figure}

~
\newline
\noindent
The polycrystalline magnetic susceptibility of \Niace\ measured at $\mu_0H = 0.1$~T is shown in Fig.~\ref{ace_data}(a) and resembles that of an $S = 1$ anisotropic magnet with negligible interactions between the spins. Similar to the previous case, the data are fitted to $\chi_{\rm av} = \frac{1}{3}(\chi_x +\chi_y + \chi_z)$ using the expressions in Eq.~\ref{sus}, with an isotropic $g$ and under the constraint $0<3E<|D|$. The fitted line reproduces the data well and yields the parameters $g = 2.20(1)$, $D = -5.7(3)$~K and $E = 1.36(3)$~K.\footnote{Note, without the $3E<|D|$ constraint, the fit also works well with $g = 2.20(1)$, $D = 4.9(2)$~K and $E = 2.2(2)$~K. This is accounted for by the reverse permutation of crystallographic axes described earlier and the parameters can be mapped back on to those above via the inverse relations of Eq.~\ref{convert}.}  

\vspace{0.3cm}
The magnetization data measured up to 7~T at various temperatures are shown in Fig.~\ref{ace_data}(b). All traces show a smooth rise towards saturation, with the 0.6~K data approaching a moment of 2.12$\mu_{\rm B}$ per Ni(II) ion by 7~T. There is no clear sign of any feature due a ground state energy level crossing either in $M(H)$ or ${\rm d}M/{\rm d}H$ (inset) at the lowest temperatures. This is consistent with the expectation of an easy-axis system. Although, as mentioned earlier, a level crossing occurs for fields parallel to $x$ for easy-axis materials, a feature in the polycrystalline magnetization data is only expected to be observed for $k_{\rm B}T\lesssim 0.1\times\sqrt{2(E^2-DE)}$. The estimate of $D$ and $E$ obtained from susceptibility suggests this condition is not met in our measurements. 

\vspace{0.3cm}
Figure~\ref{ace_data}(c) shows the zero-field heat capacity of polycrystalline \Niace. On cooling, the data exhibit a broad hump between 40 and 50~K due to phonons followed by a steep rise at low temperatures caused by single-ion anisotropy. To extract estimates of the anisotropy parameters it is necessary to fit the data below 18~K to the sum of a Debye phonon mode~\cite{SI} and the magnetic term given in Equation~\ref{zf_hc}. The resulting fit is displayed as a solid red line in the figure and is seen to compare well with the data at low temperatures. The separate Debye and magnetic terms in the fit are shown as dashed green and dotted blues lines, respectively. The anisotropy parameters resulting from the fit are easy-axis $D = -6.7(1)$~K and $E = 1.54(1)$~K,\footnote{The fit of the lattice contribution yields the following characteristic amplitude and Debye temperature, $A_{\rm D} = 79(1)$~J\,K$^{-1}$\,mol$^{-1}$, $\theta_{\rm D} = 90(1)$~K.} and are within 10---15\% of the values obtained from fitting the magnetic susceptibility.  

Additional low-temperature heat capacity measurements are made in fixed magnetic fields. The fitted zero-field lattice term is subtracted from these data and the results are plotted as $C_{\rm mag}(T)$ in Figure~\ref{ace_data}(d). In small applied fields the data exhibit the low-temperature rise due to the anisotropy. At higher fields this feature moves to higher temperatures and reveals itself to be a peak whose amplitude, width and position increase with increasing fields. This is consistent with the simulated data shown in Fig.~\ref{hcap_sim}(a), further confirming the easy-axis nature of this material.  

\subsubsection{Discussion}~
\newline
\noindent
Low-field magnetic susceptibility, magnetization and heat capacity measurements all indicate the presence of easy-axis anisotropy in \Niace\ where the Ni(II) ion is surrounded by four equatorial oxygens and two axial nitrogens. Judging from the local structure, it would be expected that the easy axis lies parallel or close to the axial N---Ni---N bond direction. The parameters taken from the heat capacity analysis, which on the evidence of the previous material offers the most accurate results, are $D = -6.7(1)$~K and $E = 1.54(1)$~K. In contrast, \Nilut, where Ni(II) is surrounded by four equatorial nitrogens and two axial oxygens, is an easy-plane system with $D = 10.4(1)$~K and $E = 2.6(2)$~K (also from heat capacity).

\section{Systems with significant exchange}

As detailed above, it is possible from polycrystalline thermodynamic measurements alone to obtain good estimates for the parameters governing the magnetic properties of $S=1$ systems in the absence of effective exchange pathways. Now we turn to systems containing antiferromagnetic interactions between the spins. The Hamiltonian in this case is 
\begin{equation}
\hat{\mathcal{H}} = \sum_{\langle i, j \rangle}J_{ij}\hat{\bf S}_i\cdot\hat{\bf S}_j + D\sum_{i}(\hat{S}^z_i)^2 + E\sum_{i}[(\hat{S}^x_i)^2  - (\hat{S}^y_i)^2]
+ \mu _{\textsc{b}}\sum_i{\bf B}\cdot{\bf g}\cdot\hat{\bf S}_i, 
\label{ham2}
\end{equation}
where the sum in the first term is over unique nearest-neighbour exchange pathways with Heisenberg exchange strength $J_{ij}$. In the two extreme cases, where the exchange term is much stronger than the anisotropy term or vice versa, then polycrystalline data can be used to parameterize the system. However, in the case where the two are similar in size then interpretation of the data can be problematic, as some of the present authors have discussed previously~\cite{NiSbF6,JamieBSbF6}. 

In this situation, there is a paucity of theoretical models that can be used in fitting either low-field magnetic susceptibility or zero-field heat capacity to obtain reliable estimates of the magnetic parameters. To some extent the application of magnetic field can help. The field both suppresses antiferromagnetism and shifts the features in heat capacity due to the anisotropy to higher temperatures, permitting them to be analysed. Features that provide useful information can also be discerned in polycrystalline measurements of $M(H)$, specifically the spin-flop field (in easy-axis systems) and the saturation fields in the easy and hard directions. Here we illustrate these methods using experimental data collected on a Ni(II) coordination polymer.

\subsection{Experimental results for \Nipyz}

  \begin{figure}[t]
\centering
  \includegraphics[width=10.0cm]{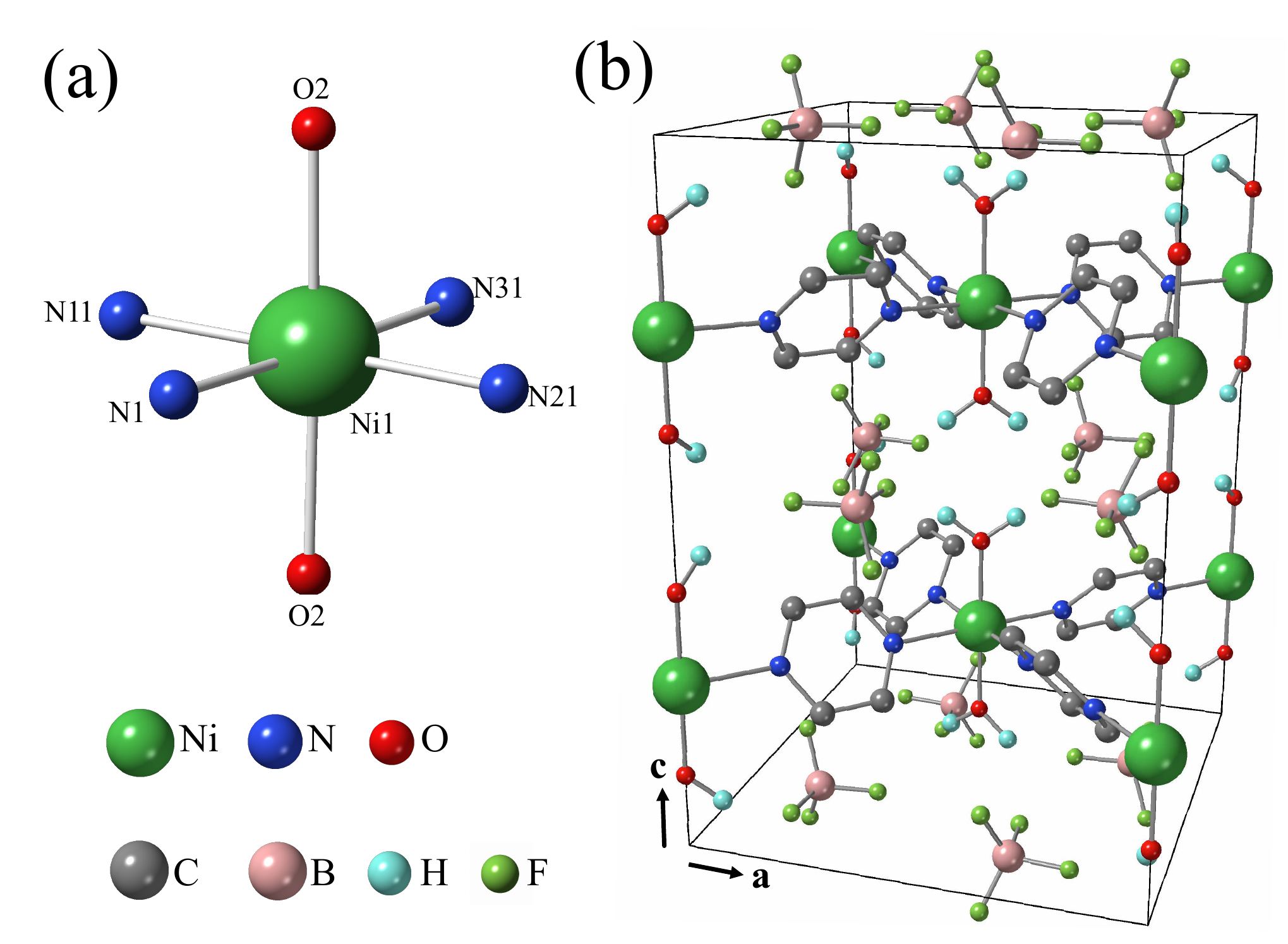}
  \caption{Room-temperature crystal structure of \Nipyz\ determined using powder synchrotron x-ray diffraction. (a) Local Ni(II) environment and atomic labelling scheme. (b) Unit cell showing Ni-pyrazine square lattices sheets. Water hydrogens can occupy four equally probable positions, one of which is shown here. Pyrazine hydrogen are omitted for clarity.} 
\label{pyz_struc}
\end{figure}

\subsubsection{Crystal structure}~
\newline
\noindent
\Nipyz~crystallizes in the tetragonal space group $I4/mcm$. Fig.~\ref{pyz_struc} shows the structure of this material as determined at 300~K using powder synchrotron x-ray diffraction. The coordination environment consists of NiN$_{4}$O$_{2}$ octahedra with a small axial compression, but no distortion in the octahedral bond angles. The equatorial nitrogens are from the pyrazine molecules, which bridge the Ni(II) ions in the $ab$-plane forming a square planar array. The axial oxygens are provided by the water molecules that tether adjacent nickel-pyrazine sheets along $c$ via a network of H$\cdots$F bonds with the charge-balancing BF$^-_4$ counter ions. 

The nearest neighbour Ni$\cdots$Ni distance is 6.98~\AA~through the pyrazine molecules, and metal ions in adjacent planes are separated by 7.40~\AA. Both pyrazine and H$\cdots$F bonds have been shown to be mediators of antiferromagnetic exchange strengths of the order of 1---10~K in Ni(II) complexes~\cite{JamieBSbF6,Junjie}. It is not possible to tell from the structure alone which pathways will support significant exchange interactions, therefore we define the average nearest-neighbour exchange strength, $\langle J \rangle$, as a sum of the exchange through pyrazine, $J_{\rm pyz}$, and water, $J_{\rm H_2O}$, such that $n\langle J \rangle = 4J_{\rm pyz} + 2J_{\rm H_2O}$, where $n$ is the total number of effective nearest-neighbour exchange pathways.

\vspace{0.3cm}
The comparison of structural parameters in Table~\ref{tab_struc} can be used to judge the extent to which the \Nilut\ system discussed earlier can be considered an exchange-free analogue of \Nipyz, as was anticipated at the design stage. While both have NiN$_{4}$O$_{2}$ coordination environments and overall the Ni-ligand distances are comparable, \Nipyz\ has four equal equatorial bond lengths and ideal octahedral bond angles, whereas \Nilut\ has four distinct equatorial bond lengths and bond angles that depart somwhat from octahedral symmetry. Thus, in contrast to the Ni-lutidine system, $E$ is expected to be zero in the high-symmetry Ni-pyrazine material. Nevertheless, the comparison suggests that $D$ should have the same sign in the two systems with a similar order of magnitude, i.e. we anticipate that \Nipyz\ has easy-plane anisotropy and $D \sim 10$~K.

\subsubsection{Thermodynamic measurements}~
\begin{figure}[t]
\centering
\includegraphics[width=12.0cm]{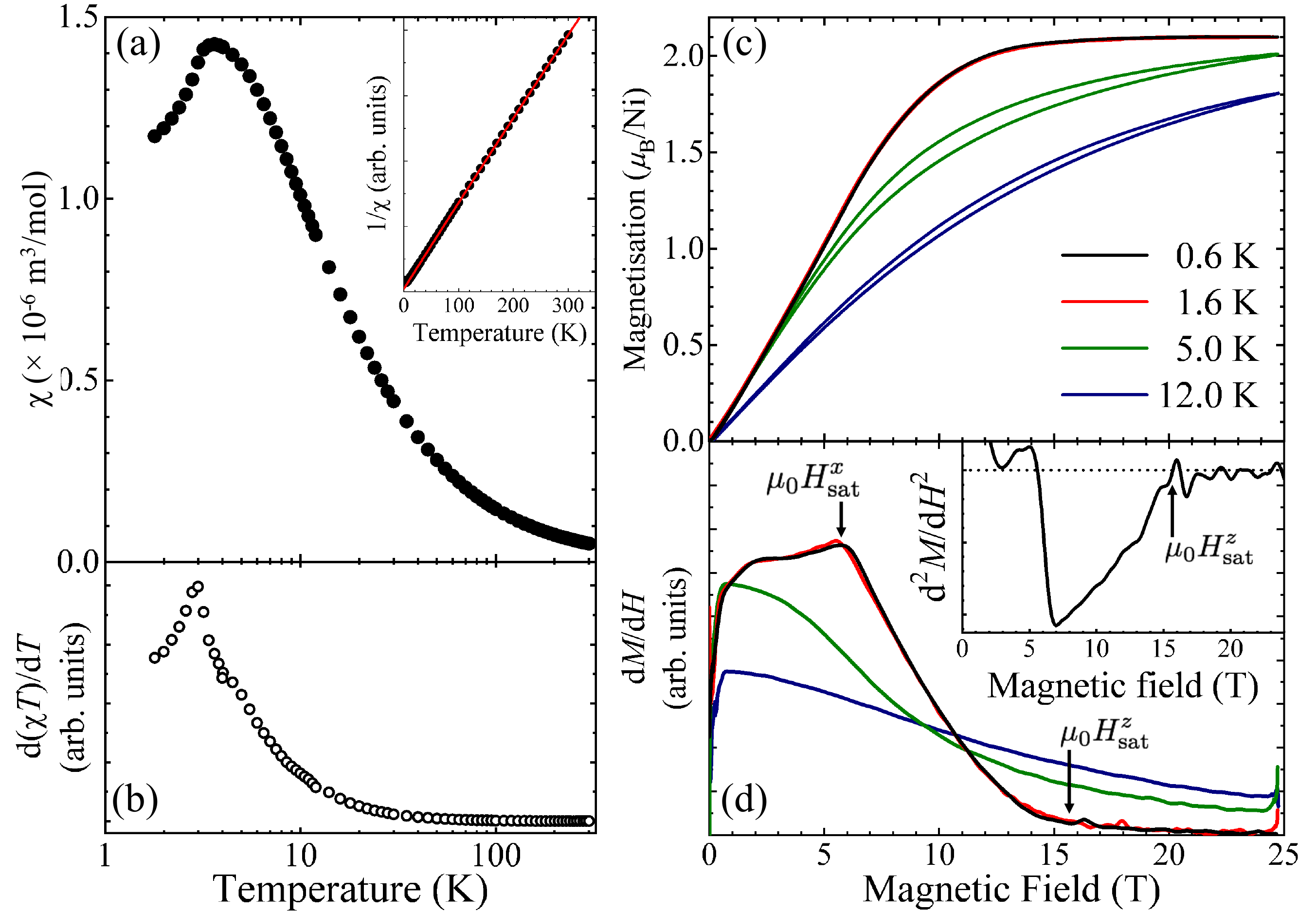}
\caption{Magnetometry data for \Nipyz. (a) Magnetic susceptibility $\chi(T)$ measured at $\mu_0H = 0.1$~T. Inset: $1/\chi(T)$ (circles) and linear fit (red line). (b) ${\rm d}(\chi T)/{\rm d}T$ exhibits a lambda-like peak at 3.0(1)~K. (c) Pulsed-field magnetization $M(H)$ measured at various temperatures. (d) ${\rm d}M/{\rm d}H$ data with the positions of the characteristic fields marked by arrows. The inset shows ${\rm d}^2M/{\rm d}H^2$ at 0.6~K. $\mu_0H^z_{\rm sat}$ is defined as the field at which the ${\rm d}^2M/{\rm d}H^2$ curve first approaches zero.} 
\label{pyz_mag}
\end{figure}
\newline
\noindent
The susceptibility of polycrystalline \Nipyz\ taken at $\mu_0H = 0.1$~T is shown in Fig.~\ref{pyz_mag}(a). The data rises smoothly on cooling and exhibits a broad maximum around 4~K, followed by a cusp and a reduction down to 1.8~K. The inverse susceptibility (inset) is fit to a Curie-Weiss model across the range $100 < T < 300$~K, yielding $g = 2.19(1)$ and a temperature-independent contribution $\chi_0 = 1.3(1)\times10^{-9}$~m$^3$mol$^{-1}$. The same data are plotted as ${\rm d}(\chi T)/{\rm d}T$ in Fig.~\ref{pyz_mag}(b). This quantity is known to resemble the behaviour of the heat capacity of simple antiferromagnets in the region of a transition to long-range order~\cite{fisher}. The data show a lambda-like peak close to the cusp observed in $\chi$, indicative of an antiferromagnetic transition at $3.0(1)$~K.

\vspace{0.3cm}
Fig.~\ref{pyz_mag}(c) shows pulsed-field magnetization data taken at various fixed values of temperature. As the field is swept, the data display a slightly concave rise followed by a rounded approach to saturation, distinctive of an $S =1$ antiferromagnet with single-ion anisotropy. Above 15~T at the lowest temperatures the moment approaches a saturated value of $2.10(1)\mu_{\rm B}$ per Ni(II), which suggests a low-temperature value of $g = 2.10(1)$. There is no indication of a spin flop in the data, which is consistent with the expectation of easy-plane anisotropy in this material. Following Reference~\cite{JamieBSbF6}, we expect to see two characteristic fields in a polycrystalline measurement of $M(H)$ of an easy-plane system: one at the point where moments saturate for fields lying in the easy-plane, and the other where moments saturate for fields parallel to the hard axis . These occur at $\mu_0H^x_{\rm sat} = 2n\langle J \rangle/g\mu_{\rm B}$ and $\mu_0H^z_{\rm sat} = 2(n\langle J \rangle + D)/g\mu_{\rm B}$, respectively. A change in curvature is observed in the low-temperature data between 5 and 6~T, which appears as a peak in ${\rm d}M/{\rm d}H$ [as shown in Fig.~\ref{pyz_mag}(d)]. We associate this feature with fields within the easy-plane and find $\mu_0H^x_{\rm sat} = 5.7(3)$~T. Furthermore, we define the hard-axis saturation as the point at which ${\rm d}^2M/{\rm d}H^2$ first approaches zero, hence $\mu_0H^z_{\rm sat} = 15.7(5)$~T. From these two values we estimate $n\langle J \rangle = 4.0(2)$~K and $D = 7.1(6)$~K.

\begin{figure}[t]
\centering
\includegraphics[width=12.0cm]{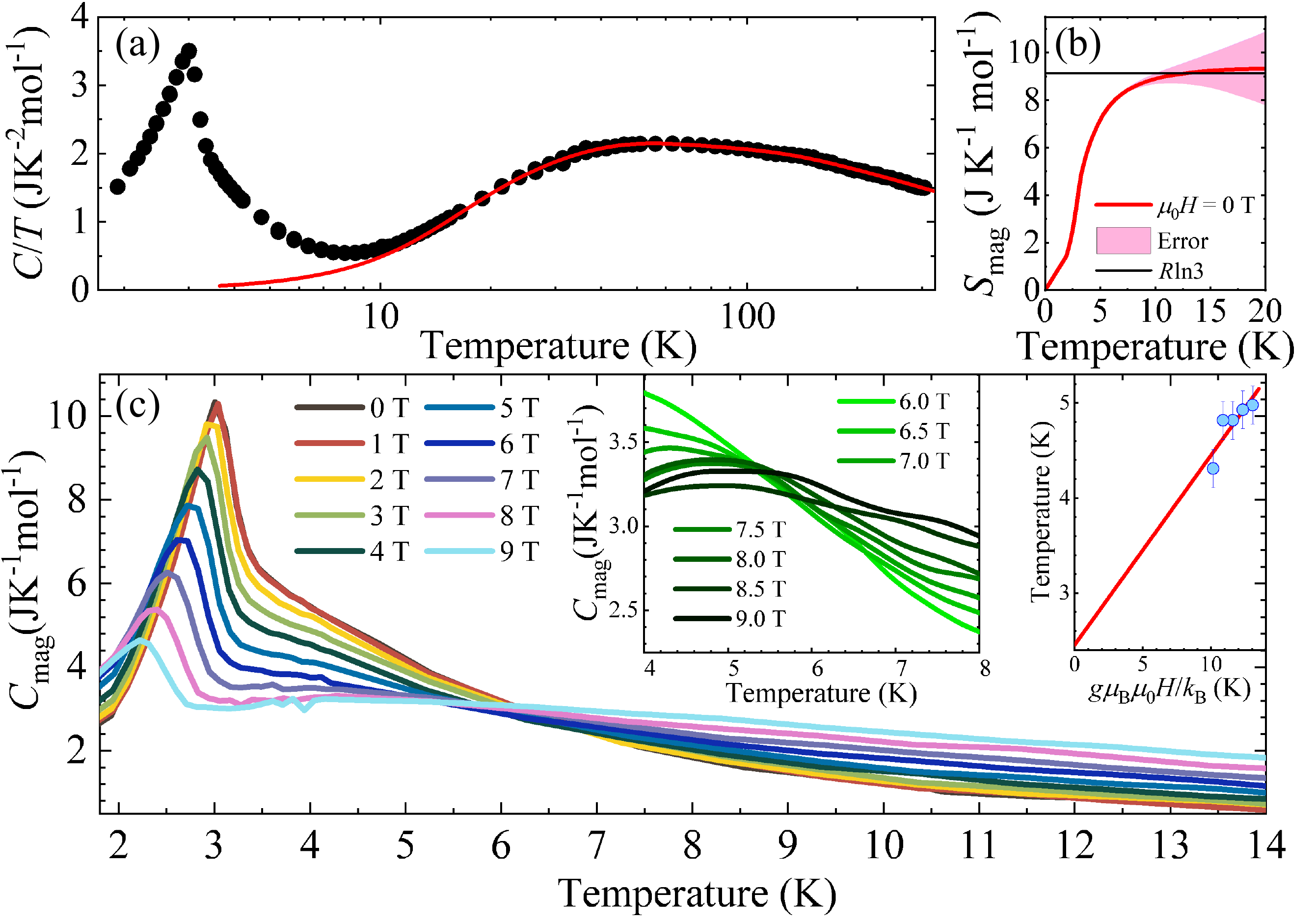}
\caption{Heat capacity measurements of \Nipyz. (a) Zero-field data plotted as $C/T$ (circles). The red line is a fit of the lattice contribution extrapolated to low temperatures. (b) Magnetic entropy up to 20~K, which approaches the expected value of $R\ln3$. (c) Magnetic heat capacity $C_{\rm mag}(T)$ at various fixed values of applied field. Left-hand inset: the region $4\le T \le 8$~K at high fields where the broad hump due to single-ion anisotropy becomes apparent. Right-hand inset: the field-dependent position of the hump (circles) and associated linear fit (red line).} 
\label{pyz_hcap}
\end{figure}

\vspace{0.3cm}
Zero-field heat capacity measurements performed on \Nipyz\ [plotted as $C/T$ in Fig.~\ref{pyz_hcap}(a)] reveal a broad hump due to phonons located around 50---70~K and a lambda peak indicating a transition to long-range antiferromagnetic order centred at $T_{\rm N} = 3.0(1)$~K, in agreement with the susceptibility value. To extract the lattice contribution, a fit is made to the data in the range $24\le T \le 300$~K using a model of one Debye and three Einstein phonon modes~\cite{SI,mansonjacs2009}.\footnote{The fit yields the following characteristic amplitudes, $A_i$~(J\,K$^{-1}$\,mol$^{-1}$), and temperatures, $\theta_i$~(K), of the Debye ($i=$~D) and Einstein ($i=$~E) phonon modes: $A_{\rm D} = 98(13)$, $\theta_{\rm D} = 116(9)$, $A_{\rm E1} = 109(6)$, $\theta_{\rm E1} = 208(21)$, $A_{\rm E2} = 234(10)$, $\theta_{\rm E2} = 509(20)$, $A_{\rm E3} = 247(27)$ and $\theta_{\rm E3} = 1287(128)$.}  The result, shown as a red line, agrees well with the data across the fitted temperature range.  The magnetic part of the heat capacity, $C_{\rm mag}$, is isolated by subtracting the lattice contribution. The magnetic entropy is calculated by integration and found to approach the expected value of $R\ln 3$ at temperatures in excess of 10~K. $C_{\rm mag}(T)$ measured in fixed applied fields is shown in Fig.~\ref{pyz_hcap}(c). The lambda peak associated with antiferromagnetic ordering is seen to be suppressed as the field is increased, while a broad shoulder appears to the high-temperature side of the peak and shifts to higher temperatures with increasing field. This broad feature is associated with the single-ion anisotropy and its temperature evolution with field is shown in more detail in the left-hand inset. The right-hand inset shows the position of the hump, which can only be discerned at the highest measured fields, plotted against $g\mu_{0}\mu_{\rm B}H/k_{\rm B}$. Following the discussion of Eq.~\ref{grad}, we perform a linear fit of these data to find $\delta = 0.20(7)$ and $\gamma |D|/k_{\rm B} = 2.5(8)$~K. Using these values together with Fig.~\ref{hcap_sim}(f), and assuming an easy-plane scenario, we estimate $D = 7(2)$~K, which agrees with the result from the magnetization data.

\subsection{Muon-spin relaxation}

In order to confirm the presence of long-range magnetic order suggested by the heat capacity data, muon-spin relaxation ($\mu^{+}$SR) measurements were made on \Nipyz. Example spectra are shown in Fig.~\ref{pyz_neut}(a). At temperatures $T<3.2$~K the asymmetry shows heavily damped oscillations at two distinct frequencies whose magnitudes decrease with increasing temperature. At temperatures $T>3.2$~K, oscillations  are seen at lower frequency, but show little variation as the temperature is further increased. The oscillations measured for $T<3.2$~K are characteristic of a quasistatic local magnetic field at the muon stopping site usually attributable to long-range magnetic order, which causes a coherent precession of the spins of those muons with a component of their spin polarization perpendicular to this local field. The frequencies of the oscillations are given by $f_{i} = \gamma_{\mu} B_{i}/2 \pi$, where $\gamma_{\mu}$ is the muon gyromagnetic ratio ($=2 \pi \times 135.5$~MHz T$^{-1}$) and $B_{i}$ is the average magnitude of the local magnetic field at the $i$th muon site. The oscillations for $T>3.2$~K are caused by dipole-dipole coupling between the muon and fluorine nuclei and are typically resolved in the paramagnetic regime. Detailed modelling of these two regimes is described in~\cite{SI}. This allows us to conclude that the material undergoes a transition to long-range order throughout its bulk at $T_{\mathrm{N}}=3.2(1)$~K, which is in excellent agreement with the heat capacity data.

\subsubsection{Neutron diffraction}~
\newline
To check the reliability of the easy-plane attribution, we performed neutron powder diffraction on a deuterated sample of [Ni(D$_2$O)$_2$(d$_4$-pyz)$_2$]($^{11}$BF$_4$)$_2$ at the WISH diffractometer (ISIS, Rutherford Appleton Laboratory, UK)~\cite{SI,WISH}. 

A full quantitative structural refinement of the data is made difficult by the dynamics of the water molecules. A LeBail fit of the nuclear Bragg peaks observed at 10~K is fully consistent with the reflection conditions of the space group $I4/mcm$ and yields lattice parameters $a = 9.8859(2)$\,\AA, $b = 9.8859(2)$\,\AA, $c = 14.6625(4)$\,\AA, which are in good agreement with the results of the structural refinement of the room-temperature x-ray diffraction data taken on the non-deuterated material. To account for the fact that the reflections with a sizeable projection on the ${\bf c^*}$ reciprocal lattice vector were found to be broader than others, the fit to the neutron data includes a strain model that represents a small degree of decoherence along the crystalline ${c}$-axis.

\begin{figure}[t]
\centering
\includegraphics[width=\textwidth]{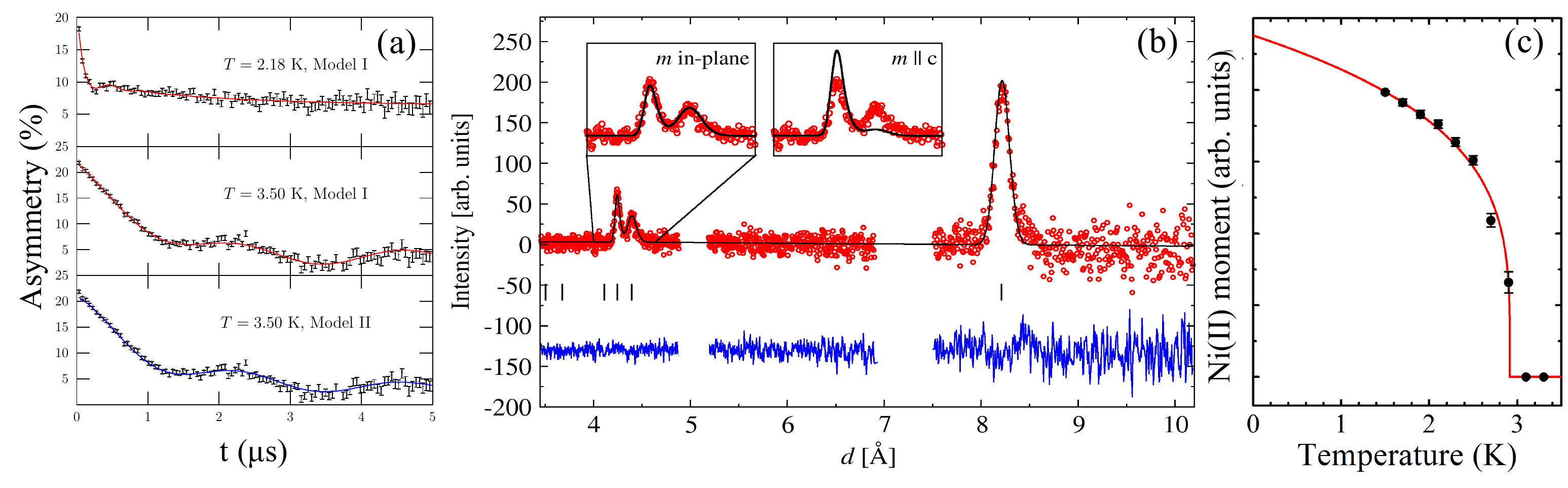}
\caption{(a) Example muon-spin relaxation spectra for \Nipyz\ at selected temperatures. Solid lines are fits described in~\cite{SI}. (b) and (c) Neutron diffraction results for [Ni(D$_2$O)$_2$(d$_4$-pyz)$_2$]($^{11}$BF$_4$)$_2$. (a) Magnetic diffraction pattern (red points) obtained by subtracting data collected at 10~K from that collected at 1.5~K~\cite{SI}. Note that the artefacts that arise in the subtraction of the brightest nuclear reflections in the presence of a slight lattice contraction have been masked. The fitted spectrum (black line) has the Ni(II) moments lying perpendicular to the $c$-axis. Bragg peaks are indicated by ticks and the blue line is the difference between the data and the fit. The insets show a comparison of the model calculated with the moments perpendicular and parallel to the $c$-axis. (b) Temperature dependence of the ordered Ni(II) magnetic moment (circles) and the power-law fit (red line) described in the text.} 
\label{pyz_neut}
\end{figure}

Taking the difference in scattered neutron intensity obtained at 1.5 and 10~K reveals three magnetic diffraction intensities~\cite{SI}, the positions of which can be indexed by the propagation vector $k=(0,0,0)$ with respect to the reciprocal lattice of the paramagnetic unit cell [see Fig.~\ref{pyz_neut}(b)]. The three peaks correspond to the following families of reciprocal lattice vectors: $\{1,0,1\}$, $\{1,0,3\}$, and $\{2,1,1\}$ and are attributed to long-range magnetic order of Ni(II) ions as observed using $\mu^+$SR.
None of the observed magnetic reflections violate the $I$-centring reflection condition, which means that magnetic moments related by $I$-centring (corresponding to a translation of $[1/2,1/2,1/2]$) must align parallel to one another. The reflections $\{1,0,1\}$ and $\{1,0,3\}$ do violate the $c$-glide reflection condition of the nuclear structure, hence magnetic moments of atoms related by the $c$-glide must align anti-parallel to one another [corresponding to a translation of $[0,0,1/2]$ for the Ni(II) sublattice]. No evidence for canted antiferromagnetism is observed in either the neutron or magnetometry data and so a collinear magnetic structure is imposed. There are four Ni(II) ions in the unit cell at positions $[0,0,1/4]$, $[0,0,3/4]$, $[1/2,1/2,1/4]$, and $[1/2,1/2,3/4]$. Thus, from the positions of the magnetic diffraction peaks alone, we can conclude that these atoms have relative magnetic moment directions up-down-down-up, respectively. Considering the full Ni(II) sublattice, this is the $G$-type magnetic structure with all nearest neighbours aligned antiferromagnetically.

The magnetic moment directions are determined by fitting the relative intensities of the magnetic diffraction peaks in the subtracted data.  Because of the difficulties with the nuclear refinement, the magnetic intensity could not be calibrated and the scale of the magnetic phase was left free to refine. As can be seen in the insets to Fig.~\ref{pyz_neut}(b), the relative diffraction intensities were consistent with magnetic moments aligned in an undetermined direction perpendicular to ${\bf c}$, confirming the presence of easy-plane anisotropy consistent with the results of thermodynamic measurements.

\vspace{0.3cm}
The square root of the integrated intensity of the $\{1,0,1\}$ magnetic peak, which is proportional to the ordered moment, is plotted as a function of temperature in Fig.~\ref{pyz_neut}(c).  A fit to a power-law dependence, $m(T)/m(0) = (1 - T/T_{\rm c})^{\beta}$, yields $T_{\rm c} = 2.91(2)$~K, which is consistent with the position of the ordering peak in the heat capacity. While the sparseness of the data in the vicinity of $T_{\rm c}$ limits the sensitivity of the fit to the exponent $\beta$, we note that the fitted value of $0.25(2)$ is in excellent agreement with values found for experimental realizations of the 2D $XY$ model~\cite{bramwell93}.

\subsection{Discussion}

Thermodynamic measurements on \Nipyz\ are all consistent with easy-plane anisotropy, which is verified by neutron diffraction. Isothermal magnetization suggests $D = 7.1(6)$~K and $n\langle J \rangle = 4J_{\rm pyz} + 2J_{\rm H_2O} = 4.0(2)$~K. The value of $D$ agrees with the analysis of the high-field heat capacity and is within 32\% of the value found above for the non-interacting system \Nilut.  Previously measured values of $J_{\rm pyz}$ in related Ni(II) compounds are $\sim 1$~K~\cite{JamieBSbF6}, which might suggest that $J_{\rm H_2O}\ll J_{\rm pyz}$ and that our material is a highly two-dimensional antiferromagnet. This would be consistent with the analysis of the critical exponent extracted from the neutron data.

\section{Using density functional theory to obtain single-ion parameters}

To model the Hamiltonian parameters in \Nipyz\ we performed a
sequence of density functional theory (DFT) total energy
calculations.
  Calculations were performed within the DFT plane wave formalism as implemented in the {\sc Castep}
  code\cite{castep1,castep2}. The exchange-correlation interactions
  were described with the PBE generalised gradient
  functional\cite{PBE}, and ultrasoft pseudopotentials\cite{USP} were
  used for the core-valence interactions. Numerical convergence of the
  plane wave basis set (plane wave cut-off and $k$-point sampling) was
  set at a tight tolerance such that total energy differences were
  converged to better than 0.01~meV/cell to obtain
  accurate results for coupling constants~\cite{ScienceDFT}. Geometry
  optimisations were performed using a BFGS energy minimisation
  algorithm until the maximim residual force on atoms were all below
  0.05~eV/\AA.

Spin-orbit coupling, implemented in {\sc Castep} with the formalism of Dal
Corso, {\it et al.}\cite{SOC} was also used, where $j=l\pm 1/2$-resolved pseudopotentials are obtained from a fully relativistic
radial atomic Dirac-like equation. This is needed because the strongest part
of the spin-orbit interaction is within the core and so, in a
plane-wave calculation, it must be dealt with via the construction of a
$j$-dependent pseudopotential. We then use a 4-component spinor as a
pseudowavefunction, rather then the usual 2 component spin up/down
formalism. The 4-component wavefunction allows for local spin
orientations and permits inclusion of spin-orbit coupling, which is closely related to
non-collinear magnetism. At each point in space there is a local
direction to the spin polarisation and this is used to
evaluate the exchange-correlation interaction using standard
functionals. The magnetic structure is not the same as the
crystallographic structure and hence in the electronic structure
calculations we do not impose a predetermined symmetry on the
electronic charge densities.

\begin{figure}[t]
\centering
\includegraphics[height=8cm]{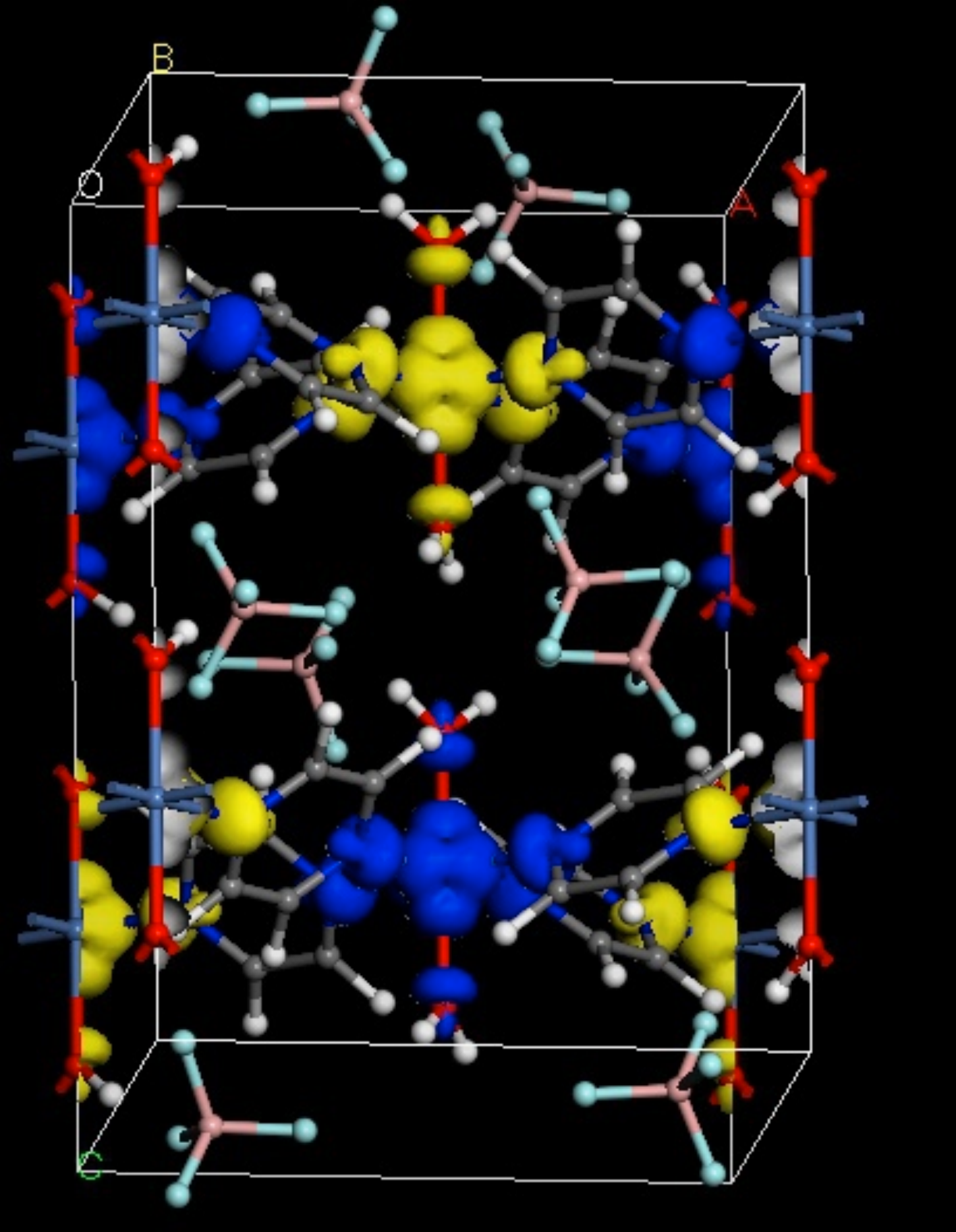}
\caption{Spin density of the lowest energy spin state of \Nipyz\, obtained from
  the density functional calculations. The blue spin
  isosurface shows one spin channel while yellow shows the other.}
\label{spin_structure}
\end{figure}

To extract the Heisenberg coupling constants, 
we employ the method described in~\cite{SI} involving comparing several
collinear spin configurations. 
After geometry optimisation of the system, there are two different pairs of
Ni-Ni interactions. The structure therefore suggests two exchange constants: $J_{1}$ within the
Ni-pyz planes and $J_{2}$ between them. In spin-polarised systems each
energy minimum in an electronic structure calculation corresponds to a
magnetic structure. To investigate likely magnetic structures
(collinear in the first instance) the electronic structure is
initialized with various spin configurations and energy minimized to
the nearest local minimum electronic magnetic state. The spin
structure of the lowest energy state is shown in
Fig.~\ref{spin_structure}. It forms an antiferromagnetic state within each
of the Ni-pyz planes, offset by (1/2, 1/2) in neighbouring planes. To evaluate the
intraplane $J_1$ coupling and the interplane (next nearest
neighbour) $J_2$ coupling, differences in energy of spin
configurations are taken giving $J_1=0.64(1)$~meV and
$J_2=0.65(1)$~meV, where the uncertainty is that of numerical noise
in the calculation\cite{JNoise}.  Our calculations therefore predict an
antiferromagnetically ordered ground state with an isotropic exchange
$J_{1}\approx J_{2}\approx 8$~K.
The calculations overestimate the
$J$-couplings compared to the experimental results. This is likely
due to the use of the PBE functional which may underestimate the
localisation of the Ni $d$ electrons, allowing slightly more
neighbour-neighbour overlap and increasing the apparent strength of the
magnetic coupling. (Such a systematic effect has been noted previously in
GGA+U calculation in
Ni-pyz-based systems \cite{NiSbF6}.) It is also worth noting that in 
previous calculations of exchange effects in a coordination polymer magnet \cite{kagome}, we found that
a similar overestimate resulted from the neglect in the calculations of
the effects of structural disorder, which acted to strongly reduce the
exchange coupling. As noted above (Fig.~\ref{pyz_struc}), the water molecules that mediate the interplane exchange in this material exhibit a degree of positional disorder. This then could act to suppress the interplane coupling, leading to a quasi-two-dimensional magnet, which would be consistent with the critical exponent extracted from the neutron data.

In addition to exchange, the energy scales of single-ion anisotropy effects can be investigated by examining the
dependence of the energy on the direction of the spin configurations.
This requires that the spin-orbit interaction is taken into account in
the electronic structure calculations and that the spins are allowed to adopt non-collinear
configurations. 
  Spin anisotropy, $D$, of the system can be found by examining energy
differences for the spin configurations that are possible in various
orientations. For this the atomic anisotropic energy expression $H  =  D S_x^2 + E\left(S_x^2-S_y^2\right)$ is used. Electronic structure calculations are carried out,
initialising the spins to be aligned along the $x$, $y$ and $z$
directions. We find $D=8.5(2)$~K and $E \leq 0.2$~K. The prediction for $D$ is in good agreement with the value of 7.1(6)~K established using magnetometry above. The value of $E$ falls within the limits of resolution of the calculation itself\cite{JNoise}, and so can be considered zero within the errors, which agrees with the expectation that $E = 0$ in this tetragonal system.
Lastly, we note that, in computing these values, it is important
to include spin-orbit coupling since this  contributes significantly to the
anisotropy coefficient. Similar calculations without spin-orbit
coupling greatly underestimate $D$, giving $|D|\lesssim 0.2$~K.

\section{Conclusions}

In this paper we have presented an experimental method for extracting the anisotropy parameters of polycrystalline $S = 1$ magnets from thermodynamic data and applied it to the situation of magnetically-isolated, exchange-free systems as well as an extended material with antiferromagnetic exchange pathways. We have sought to determine to what extent quantitative information can be achieved using readily accessible measurement techniques. 

\Nilut\ was shown to be an exchange-free, easy-plane system with $D = 10.4(1)$~K and $E = 2.6(2)$~K with coordination environment NiN$_4$O$_2$. In contrast, \Niace\ has environment NiO$_4$N$_2$ and is easy axis with parameters $D = -6.7(1)$~K and $E = 1.54(1)$~K. Based on the experimental data, \Nipyz\ is an easy-plane antiferromagnet formed from square planar Ni---pyrazine sheets separated by H$_2$O and BF$^-_4$ molecules, with $D = 7.1(6)$~K, $E = 0$ and $n\langle J \rangle = 4.0(2)$~K, where $n$ is the number of active exchange pathways. The system orders antiferromagnetically below 3.2(1)~K.

More generally, in the case of the exchange-free magnets, we have found that fitting zero-field heat capacity data yields reliable values for $D$ and $E$, and confirmed this using high-frequency ESR. The field-dependence of the heat capacity is a useful check on the sign of the $D$ and the magnitudes of the parameters can be further confirmed by features in the isothermal magnetization. Low-field magnetic susceptibility is a relatively quick and available technique. We find that fitting the results of such measurements using the expressions described can provide rough estimates of $D$ and $E$. For the antiferromagnetic system we were unable to extract quantitative information about the anisotropy from low-field susceptibility and heat capacity data. Instead, low-temperature magnetization in fields up to the hard-axis saturation is required to find this information. The results can be checked by measuring heat capacity in fields sufficiently high to separate the antiferromagnetic ordering peak and the anomalies that arise due to energy level splittings. 

In all cases, the experiments require temperatures low compared to the anisotropy energy. For the exchange-free systems, successful magnetization and fixed-field heat capacity measurements also depend upon applying magnetic fields which are on the scale of the anisotropy energy. The values of anisotropy found above for our materials are representative of those in octahedrally coordinated Ni(II) compounds~\cite{boca}, suggesting that the measurements can be performed in standard, commercially available equipment. On the other hand, the magnetization measurements needed to find useful information in the case of the antiferromagnet materials requires fields $\sim (n\langle J \rangle + D)/\mu_{\rm B}$. In the case of \Nipyz, $\langle J \rangle$ is small and so fields $< 16$~T were sufficient. In other molecule-based Ni(II) systems, fields in the range of pulsed magnets ($\sim70$~T) may be required. 

For the six-coordinate Ni(II) complexes described here, we have observed a correlation between the Pauling electronegativity (EN) value of the ligand donor atoms and the magnetic ground state. In the case of \Nilut\ and \Niace, {\it trans}\,-NiO$_2$N$_4$ and NiO$_4$N$_2$ octahedra are found for which N and O donor atoms have EN values of 3.04 and 3.44, respectively. The difference in EN values determines the resultant Ni(II) magnetic moment direction. Thus, the Ni(II) moment lies in the direction that includes the donor atoms of lower EN; i.e., either along the N---Ni---N axis (Ising-like) or within the NiN$_4$ plane ($XY$-like). This observation is fully consistent with all of the thermodynamic data. 

Finally, we have described density functional theory calculations that incorporate spin-orbit coupling in order to estimate single-ion anisotropy parameters. The calculated values for \Nipyz\ agree well with the experimentally derived results. Further calculations on other materials are underway to verify the wider applicability of this approach.

\section{Acknowledgements}

PAG acknowledges that this project has received funding from the European Research Council (ERC) under the European Union's Horizon 2020 research and innovation programme (Grant Agreement No. 681260). 
This work is also supported by EPSRC (including EP/N023803/1 and EP/N024028/1). 
We acknowledge Durham for Hamilton supercomputer time and also EPSRC Grant EP/P022782/1 for Archer compute resource. 
RDJ acknowledges financial support from the Royal Society. 
A portion of this work was performed at the National High Magnetic Field Laboratory, which is supported by National Science Foundation Cooperative Agreement No. DMR-1157490 and the State of Florida, as well as the  {\it Strongly Correlated Magnets} thrust of the DoE BES ``Science in 100 T'' program. 
JAS acknowledges support from the Independent Research/Development (IRD) program while serving at the National Science Foundation.
We are grateful to S$\mu$S for the provision of beamtime and to A. Amato and H. Luetkens for experimental assistance. 
NSF's ChemMatCARS Sector 15 is principally supported by the Divisions of Chemistry (CHE) and Materials Research (DMR), National Science Foundation, under grant number NSF/CHE-1346572.  
Use of the Advanced Photon Source, an Office of Science User Facility operated for the U.S. Department of Energy (DOE) Office of Science by Argonne National Laboratory, was supported by the U.S. DOE under Contract No. DE-AC02-06CH11357. 
FX would like to acknowledge the funding from the European Union's Horizon 2020 research and innovation program under the Marie Sk\l{}odowska-Curie grant agreement No 701647. 
The work at EWU was supported by the NSF through grant no. DMR-1703003. 
Data presented in this paper resulting from the UK effort will be made available at XXXXXXX.

\renewcommand{\theequation}{A\arabic{equation}}
\setcounter{equation}{0} 
\section*{Appendix A. Single-ion anisotropy calculations} 

In zero field, the Hamiltonian of an $S = 1$ exchange free system may be written
\begin{equation}
%\begin{aligned}
\hat{\mathcal{H}} = D\hat{S}_z^2 + E(\hat{S}_x^2  - \hat{S}_y^2)\\
= \begin{pmatrix}
D & 0 & E \\
0 & 0 & 0 \\
E & 0 & D \\
\end{pmatrix},
%\end{aligned}
\label{ham1}
\end{equation}
where we have used the $S=1$ spin matrices. 
As mentioned in the manuscript, equivalent ways to write this Hamiltonian can be obtained by permutation of the coordinate axes. For example, we can make the transformation $(x,y,z)\rightarrow (z',x',y')$. 
\begin{equation}
\begin{aligned}
\hat{\mathcal{H}_2} = D_1\hat{S}_{y'}^2 + E_1(\hat{S}_{z'}^2-\hat{S}_{x'}^2)
= \begin{pmatrix}
\frac{1}{2}[D_1+E_1] & 0 & -\frac{1}{2}[D_1+ E_1] \\
0 & D_1-E_1 & 0 \\
-\frac{1}{2}[D_1+ E_1] & 0 & \frac{1}{2}[D_1+E_1] \\
\end{pmatrix}
\end{aligned}
\end{equation} 
In a powder measurement, all information about the identity of $z$ with respect to the crystallographic axes is lost. In this case we can shift the Hamiltonian by a constant energy and write it in the standard form,
\begin{equation}
\begin{aligned}
\hat{\mathcal{H}_2} = \begin{pmatrix}
D_2 & 0 & E_2 \\
0 & 0 & 0 \\
E_2 & 0 & D_2 \\
\end{pmatrix}+(D_1-E_1)\mathcal{I},
\end{aligned}
\label{ham2}
\end{equation} 
where $\mathcal{I}$ is the $3\times 3$ identity matrix. The two sets of anisotropy parameters can thus be interconverted via the relations 
\begin{equation}
\begin{aligned}
D_2 &= \frac{1}{2}(3E_1-D_1)\\
E_2 &=  -\frac{1}{2}(D_1+E_1).
\end{aligned}
\label{convert}
\end{equation}
Only one set of parameters will fulfil the constraint $0< 3E_i < |D_i|$.

\vspace{0.3cm}
The other cyclic permutation $(x,y,z)\rightarrow (y',z',x')$ is also possible. In this case the Hamiltonian must be shifted by $(D_{1}+E_{1})$ and the transformed parameters are given by
\begin{equation}
\begin{aligned}  
D_{3} &= -\frac{1}{2}(D_1+ 3E_{1})\\
E_{3} &= \frac{1}{2}(D_{1}-E_{1}).
\end{aligned}
\end{equation}

We note that the full derivation of the Hamiltonian (Equation~\ref{ham}) treats the spin-orbit and Zeeman interactions as perturbations to the ground state of a single
ion \cite{yosida}. This leads to an expression for the anisotropy Hamiltonian written in terms of the matrix elements of the orbital angular momentum operator
$\hat{L}^{\mu}$. We define
\begin{equation}
\Lambda^{\mu} = \sum_{n}\frac{\langle 0 |\hat{L^{\mu}}|n\rangle\langle n |\hat{L}^{\mu}|0\rangle}{E_{n}-E_{0}},
\end{equation}
where $|0\rangle$ is the unperturbed ground state of the system with energy $E_{0}$ and $|n\rangle$ are the excited states with energies $E_{n}$. The anisotropy Hamiltonian is then given by
\begin{equation}
 \frac{ \hat{\mathcal{H}}}{(-\lambda)} =
\left[\Lambda^{z} - \frac{1}{2}(\Lambda^{x}+\Lambda^{y})\right]
(\hat{S}^z)^2
+\frac{1}{2}(\Lambda^{x}-\Lambda^{y})
[(\hat{S}^{x})^2- (\hat{S}^{y})^2]
+(\Lambda^{x}+\Lambda^{y}),
\end{equation}
where $\lambda$ is the spin-orbit interaction constant. On making the permutations of the coordinate axes it can be verified that the original matrix elements are recovered by making the substitutions given above. 

\renewcommand{\theequation}{B\arabic{equation}}
\setcounter{equation}{0} 
\section*{Appendix B. Curie-Weiss temperatures}

The expressions for the magnetic susceptibility along the principal axes of the exchange-free $S = 1$ system are given in Equation~\ref{sus}. Evaluating the expressions in the limit $E\to0$ and $g_x=g_y=g_z=g$ yields
\begin{equation}
\begin{aligned}
\chi_{xy} &= \frac{2N_{\rm A}\mu_0g^2\mu_{\rm B}^2}{D}~\frac{1- {\rm e}^{-\beta D}}{1+2{\rm e}^{-\beta D}}\\  
\chi_z &= \frac{2N_{\rm A}\mu_0g^2\mu_{\rm B}^2}{k_{\rm B}T}~\frac{{\rm e}^{-\beta D}}{1+2{\rm e}^{-\beta D}},
\label{susDonly}
\end{aligned}
\end{equation} 
as found elsewhere (e.g.~\cite{kahn}). Inverting these expressions and expanding in the limit of $k_{\rm B}T\gg D$ gives the high-temperature susceptibility in the planar and axial directions:
\begin{equation}
\begin{aligned}
{\chi_{xy}} &\approx \frac{2N_{\rm A}\mu_0g^2\mu_{\rm B}^2}{3k_{\rm B}}\left(\frac{1}{T-\frac{1}{6}D}\right)\\  
{\chi_z} &\approx \frac{2N_{\rm A}\mu_0g^2\mu_{\rm B}^2}{3k_{\rm B}}\left(\frac{1}{T+\frac{1}{3}D}\right),
\label{susDonly}
\end{aligned}
\end{equation} 
which resemble a Curie-Weiss behaviour with apparent Weiss temperatures $\Theta_{xy}\approx D/6$ and $\Theta_z\approx -D/3$, respectively. Taking a simple approximation to a powder average $\chi_{\rm av} = \frac{1}{3}(\chi_x +\chi_y + \chi_z)$ of these expressions again yields a Curie-Weiss form in the high-temperature limit, but with a zero Weiss temperature. The simulations shown in Figure~\ref{sus_fig} also indicate that $\theta_{\rm av}\approx 0$ for systems with non-zero $E$.

\section{References}
\bibliography{NiPowderRef}

\end{document}

\title[Supplementary Material]{Supplementary Material Accompanying\\
{\it ``Determining the anisotropy and exchange parameters of polycrystalline spin-1 magnets''}}

\author{W. J. A. Blackmore$^1$,
J. Brambleby$^1$,
T. Lancaster$^{2}$,
S. J. Clark$^{2}$,
R. D. Johnson$^3$,
J. Singleton$^4$,
A.~Ozarowski$^5$,
J. A. Schlueter$^{6,7}$,
Y.-S. Chen$^8$,
A. M. Arif\,$^9$,
S. Lapidus$^{10}$,
F. Xiao$^{11,12}$,
R. C. Williams$^1$, 
S. J. Blundell$^{3}$,
M. J. Pearce$^1$,
M. R. Lees $^1$,
P. Manuel$^{13}$,
D. Y. Villa$^{14}$,
J. A. Villa$^{14}$,
J. L. Manson$^{14}$ and
P. A. Goddard$^1$}

\address{$^1$Department of Physics, University of Warwick, Coventry, CV4 7AL, UK}
\address{$^{2}$Centre for Materials Physics, Durham University, Durham, DH1 3LE, UK}
\address{$^{3}$Clarendon Laboratory, University of Oxford, Parks Road, Oxford OX1 3PU, UK}
\address{$^4$NHMFL, Los Alamos National Laboratory, Los Alamos, NM 87545, USA}
\address{$^5$NHMFL, Florida State University, Tallahassee, FL 32310, USA}
\address{$^6$Materials Science Division, Argonne National Laboratory, IL 60439, USA}
\address{$^7$Division of Materials Research, National Science Foundation, 2415 Eisenhower Ave, Alexandria, VA 22314, USA}
\address{$^8$ChemMatCARS, Advanced Photon Source, Argonne National Laboratory, IL 60439, USA}
\address{$^9$Department of Chemistry, University of Utah, Salt Lake City, UT 84112, USA}
\address{$^{10}$X-ray Sciences Division, Argonne National Laboratory, IL 60439, USA}
\address{$^{11}$Department of Chemistry and Biochemistry, University of Bern, 3012 Bern, Switzerland}
\address{$^{12}$Laboratory for Neutron Scattering and Imaging, Paul Scherrer Institut, CH-5232 Villigen PSI, Switzerland}
\address{$^{13}$ISIS Facility, Rutherford Appleton Laboratory, Chilton, Didcot, Oxon, OX11 0QX, UK}
\address{$^{14}$Department of Chemistry and Biochemistry, Eastern Washington State University, Cheney, WA 99004, USA}
\eads{\mailto{jmanson@ewu.edu}, \mailto{p.goddard@warwick.ac.uk}}

\section{Experimental Details}

\subsection{Synthesis and crystal structure}

\subsubsection{Synthesis of \Nilut}~

\noindent 
Ni(BF$_{4}$)$_{2}\cdot6$H$_{2}$O (0.2543 g, 0.75 mmol) was dissolved in 1~mL of H$_{2}$O and 3~mL of 3,5-lutidine was layered on top. Upon slow reaction at room temperature over a period 
of 2 weeks small blue crystals were obtained in good yield.

\subsubsection{Synthesis of \Niace}~

\noindent
Nickel(II) acetate tetrahydrate (0.2506 g, 1.00 mmol) was dissolved in 2~mL of distilled H$_2$O to afford a blue-green solution. To this solution was added 5~mL of neat 4-picoline to give a deep blue reaction mixture. No precipitate formed during initial mixing. Upon standing at room temperature for about 6 weeks, small dark blue crystals formed which were carefully collected by suction filtration (0.3247 g). The crystals were stored in a refrigerator to preserve their quality. The substance formed was identified as Ni(acetate)$_2$(H$_2$O)$_2$(4-picoline)$_2$ by single crystal x-ray diffraction. 

\subsubsection{Synthesis of \Nipyz}~ 

\noindent 
The most efficient procedure entails a mechanochemical process. Ni(BF$_{4}$)$_{2}\cdot6$H$_{2}$O (0.4006 g, 1.18 mmol) was mixed with neat pyrazine (0.1882 g, 2.34 mmol) and the reagents ground together using a mortar and pestle. During the grinding procedure, a moist blue solid was obtained, transferred to a stainless steel autoclave, and placed in an oven set at 110\,$^{\circ}$C. The reaction mixture was held at 110\,$^{\circ}$C for 3 days and then removed from the oven to slowly cool back to room temperature after which the vessel was opened. A homogeneous pale purple powder was obtained. Under magnification, no indication for unreacted starting materials was evident.

\subsection{X-ray diffraction}

\begin{figure}[t]
\centering
\includegraphics[height=6.5cm]{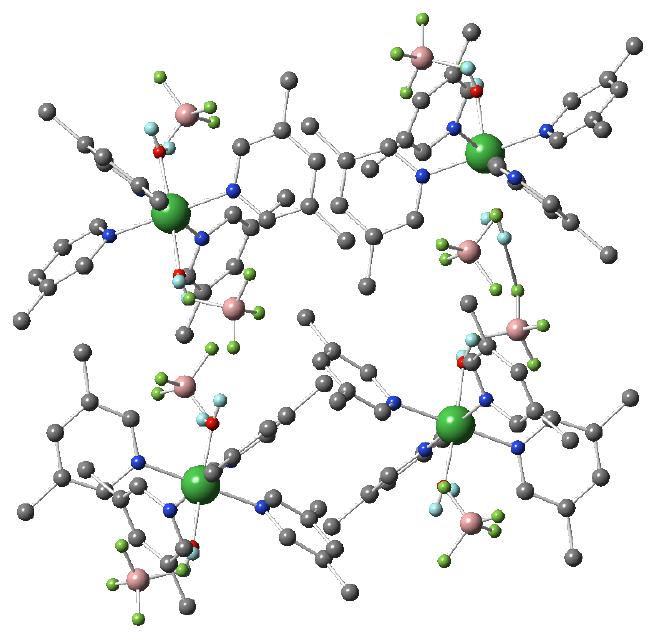}
\includegraphics[height=6.5cm]{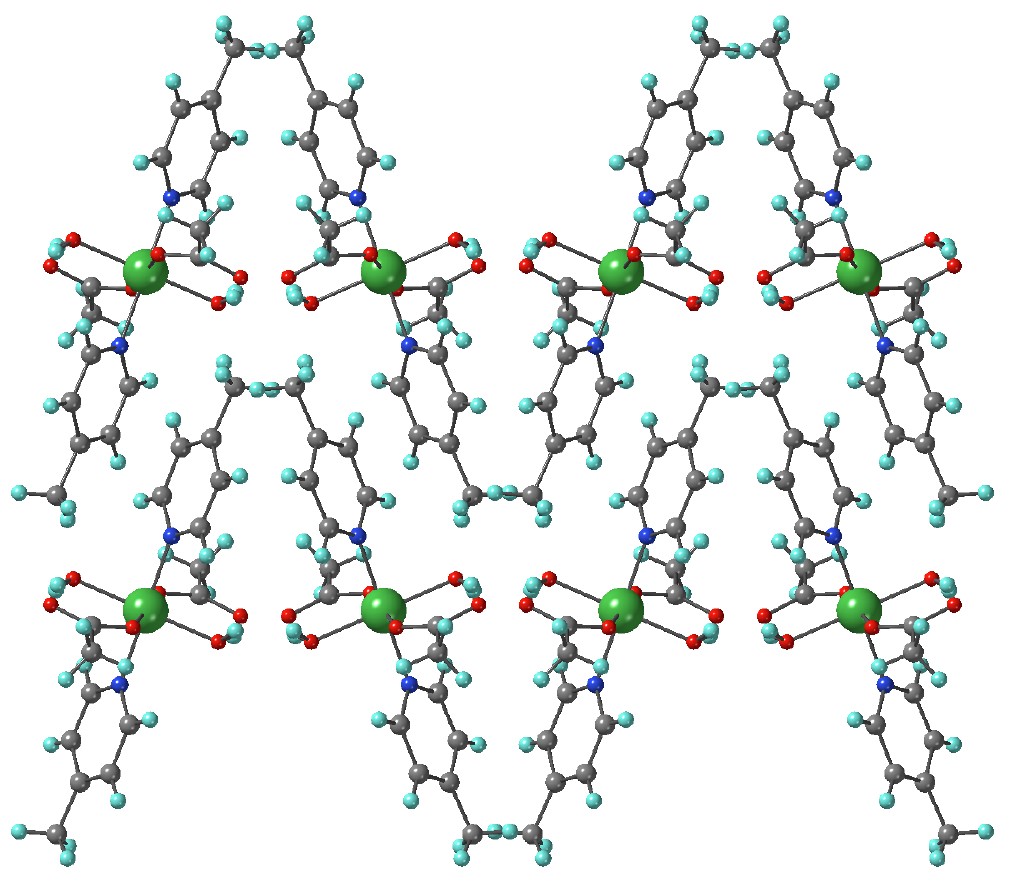}
\caption{Stacking of the molecular units in \Nilut\ (left) and \Niace\ (right). The view is along the $a$ axis in both cases. The atoms follow the same colour scheme as in Fig.~9 of the main text.} 
\label{struc_SI}
\end{figure}

\subsubsection{Single crystal x-ray diffraction of \Nilut}~ 

\noindent 
A single crystal of $\sim50\times50\times50$~$\mu$m$^3$ in size was mounted on a glass fibre through use of Paratone N oil. Data collection was carried out at the 15ID ChemMatCARS beamline of the Advanced Photon Source, Argonne National Laboratory. The data were collected at 100 K with a Bruker 6000 CCD detector. The structural refinement was carried out through use of Bruker APEX3 software v2015.5-2~\cite{apex}. All non-hydrogen atoms were refined with anisotropic displacement parameters. Hydrogen atoms, other than those associated with the water molecule, were placed in calculated positions and refined isotropically using a riding model. Hydrogen atoms associated with the water molecules were located on a difference map and their location restrained by the geometry of an isolated water molecule. The Ni(II) local environment is shown in Fig.~5 of the main text and the stacking of the molecular units is shown in Fig.~\ref{struc_SI}.

\subsubsection{Single crystal x-ray diffraction of \Niace}~

\noindent
A blue prism-shaped crystal $0.33\times0.23\times0.15$~mm$^3$ in size was mounted on a glass fibre with traces of viscous oil and then transferred to a Nonius KappaCCD diffractometer equipped with  Mo $K_\alpha$ radiation ($\lambda = 0.71073$~\AA).  Ten frames of data were collected at 150(1)~K with an oscillation range of 1~deg/frame and an exposure time of 20~sec/frame~\cite{nonius}. Indexing and unit cell refinement based on all observed reflections from those ten frames, indicated an orthorhombic $P$ lattice.  A total of 4088 reflections ($\theta_{\rm max} = 27.484^\circ$) were indexed, integrated and corrected for Lorentz, polarization and absorption effects using DENZO-SMN and SCALEPAC~\cite{otwinoski} . Post refinement of the unit cell gave $a = 8.8996(3)$~\AA, $b = 12.3995(4)$~\AA, $c = 17.6516(7)$~\AA, and $V = 1947.86(12)$~\AA$^3$.  Axial photographs and systematic absences were consistent with the compound having crystallized in the orthorhombic space group $Pcab$.

The structure was solved using SIR 97~\cite{sir}.
All of the non-hydrogen atoms were refined with anisotropic displacement coefficients. Hydrogen atoms were assigned isotropic displacement coefficients $U({\rm H}) = 1.2U({\rm C})$ or $1.5U({\rm C_{methyl}})$, and their coordinates were allowed to ride on their respective carbons using SHELXL~\cite{shelxl}. The weighting scheme employed was $w = 1/[\sigma^2(F_0^2 ) + (0.0635P)^2  + 3.6101P]$ where $P = (F_0^2 + 2F_c^2 )/3$. The refinement converged to $R1 = 0.0513$, $wR2 = 0.1356$, and $S = 1.145$ for 1515 reflections with $1> 2\sigma(I)$, and $R1 = 0.0844$, $wR2 = 0.1515$, and $S = 1.145$ for 2225 unique reflections and 121 parameters.
\footnote{$R1 = \sum( ||F_0| - |F_c || ) / \sum |F_0|$, $wR2 = \sqrt{\sum(w(F_0^2 - F_c^2)^2) / \sum(F_0^2)^2}$, and $S =$~goodness-of-fit on $F^2 = \sqrt{\sum w(F_0^2 - F_c^2)^2 / (n-p)}$, where $n$ is the number of reflections and $p$ is the number of parameters refined.} 
The maximum $\Delta/\sigma$ in the final cycle of the least-squares was 0, and the residual peaks on the final difference-Fourier map ranged from $-0.741$ to $0.951\,e/$\AA$^3$.  Scattering factors were taken from the International Tables for Crystallography, Volume C~\cite{inttab1,inttab2}. The resulting Ni(II) local environment is shown in Fig.~5 of the main text and the stacking of the molecular units is shown in Fig.~\ref{struc_SI}.

\subsubsection{Synchrotron x-ray powder diffraction of \Nipyz}~ 

\noindent 
High-resolution room temperature data were collected using beamline 11-BM at the Advanced Photon Source (APS) and 0.413841~\AA\, x-rays. Discrete detectors are scanned at a speed of 0.01$^{\circ}$/s over 34$^{\circ}$ in 2$\theta$-range with data points collected every 0.001$^{\circ}$. Data were collected while continually scanning the diffractometer 2$\theta$-arm. Peak indexing was performed using TOPAS Academic. Hydrogen atoms were placed in calculated positions using a fixed C---H bond length of 0.95 \AA. Each water hydrogen atom is positionally disordered over four equivalent sites. 

\subsection{Magnetometry}

\noindent 
Pulsed-field magnetization measurements were performed  at the National High Magnetic Field Laboratory in Los Alamos; fields of up to 40~T with typical rise times $\approx 10$~ms were used. Powdered samples are mounted in 1.3~mm diameter PCTFE ampules (inner diameter 1.0~mm) that can be moved into and out of a 1500-turn, 1.5~mm bore, 1.5~mm long compensated-coil susceptometer, constructed from 50~gauge high-purity copper wire~\cite{njp}. When the sample is within the coil and the field pulsed the voltage induced in the coil is proportional to the rate of change of magnetization with time, $({\rm d} M/{\rm d}t)$. Accurate values of the magnetization are obtained by numerical integration of the signal with respect to time, followed by subtraction of the integrated signal recorded using an empty coil under the same conditions~\cite{njp}. The magnetic field is measured via the signal induced within a coaxial 10-turn coil and calibrated via observation of de Haas--van Alphen oscillations arising from the copper coils of the susceptometer~\cite{njp}. The susceptometer is placed inside a $^3$He cryostat, which can attain temperatures as low as 500~mK.

\vspace{0.3cm}
Pulsed-field data were calibrated using measurements of the sample moment, $m$, made in a Quantum Design MPMS 7~T SQUID magnetometer fitted with DC transport mode. Powdered samples of a known mass are loaded into a gelatin capsule, placed in a plastic drinking straw and attached to a sample rod. The samples were cooled in zero-field to the measured temperature, then the field was increased in increments between $\mu_0H=0$ and 7 T and the field stabilised before each measurement was taken. The pulsed-field measurements were then scaled onto the SQUID data. The same experimental set up was used to obtain DC susceptibility measurements, with the sample zero-field cooled to 1.8~K. The field was then set to 0.1~T and temperature dependent data was taken on warming to 300~K. The molar magnetic susceptibility ($\chi_{\rm mol}$) is deduced from this measurement using $\chi_{\rm mol} = m/nH$, where $n$ is the number of moles of sample. 

\subsection{Heat capacity}

A Quantum Design 9~T PPMS instrument was used to measure temperature-dependent data in fields between 0 and 9~T. The sample was pressed into a thin pellet and mounted on a sapphire sample platform using a small amount of Apiezon-N grease to create a good thermal contact. The platform houses an electric heater and Cernox thermometer and is connected to the thermal bath by thermally conducting wires. The sample was cooled to 1.8~K in a high vacuum and the heat capacity was measured using a thermal relaxation technique. The sample platform and grease addenda was measured over the same field and temperature range beforehand and subtracted from the measurement to obtain the heat capacity of the sample. 

\vspace{0.3cm}
To extract the magnetic heat capacity, the lattice contribution is determined by fitting to the high-temperature data. We use the following model~\cite{mansonjacs2009}, which has, as adjustable parameters, the characteristic amplitudes, $A_i$~(J\,K$^{-1}$\,mol$^{-1}$), and temperatures, $\theta_i$~(K), of the Debye ($i=$~D) and Einstein ($i=$~E) phonon modes.
\begin{equation}
C_{\mathrm{latt}} = \frac{3A_{\textsc{d}}}{x^{3}_{\textsc{d}}}\int_{0}^{x_{\textsc{d}}} \!\frac{x^{4}e^{x}}{\left(e^{x}-1\right)^{2}}\,{\rm d}x + \sum_{j=1}^{n} A_{\textsc{e}j}\frac{\theta^{2}_{\textsc{e}j}}{T^{2}}\frac{e^{\left(\theta_{\textsc{e}j}/T\right)}}{\left[e^{\left(\theta_{\textsc{e}j}/T\right)}-1\right]^{2}},
\label{LatticeFit}
\end{equation} 
\noindent
where $x_{\rm D} = \theta_{\rm D}/T$. The first term is the response from a Debye mode and the second term accounts for $n$ Einstein modes.

\subsection{Electron-spin resonance}
 
High-field, high-frequency ESR spectra of powdered samples at temperatures ranging from $\approx$ 3 to 80~K were recorded on a home-built spectrometer at the EMR facility of the National High Magnetic Field Laboratory with the microwave frequencies 52--626 GHz. The instrument is a transmission-type device and uses no resonant cavity. The microwaves were generated by a phase-locked Virginia Diodes source, generating frequency of $13\pm1$~GHz, and equipped with a cascade of frequency multipliers to generate higher harmonic frequencies. A superconducting magnet (Oxford Instruments) capable of reaching a field of 17~T was employed. 

\subsection{Muon-spin relaxation}

Zero-field muon spin relaxation ($\mu^+$SR) measurements on \Nipyz\ were carried out on polycrystalline samples using the GPS spectrometer
at the Swiss Muon Source (S$\mu$S), Paul Scherrer Institut, Switzerland. For the measurements, samples were packed in Ag foil envelopes (foil thickness 25\,$\mu$m),
attached to a silver fork~\cite{steve} and mounted in a He cryostat.

In a $\mu^{+}$SR experiment \cite{steve} spin-polarized positive muons are stopped in a target sample, where the muon usually occupies an interstitial position in the crystal. The observed property in the experiment is the time evolution of the muon spin polarization, the behaviour of which depends on the
local magnetic field at the muon site. Each muon decays, with an average lifetime of 2.2~$\mu$s, into two neutrinos and a positron, the latter particle being emitted preferentially along the instantaneous direction of the muon spin. Recording the time dependence of the positron emission directions therefore allows the determination of
the spin-polarization of the ensemble of muons. In our experiments positrons are detected by detectors placed forward (F) and backward (B) of the initial muon polarization direction. Histograms $N_{\mathrm{F}}(t)$ and $N_{\mathrm{B}}(t)$ record the number of positrons detected in the two detectors as a function of time following the muon implantation. The quantity of interest is the decay positron asymmetry function, defined as
\begin{equation}
A(t)=\frac{N_{\mathrm{F}}(t)-\alpha_{\mathrm{exp}} N_{\mathrm{B}}(t)}
{N_{\mathrm{F}}(t)+\alpha_{\mathrm{exp}} N_{\mathrm{B}}(t)} \, ,
\end{equation}
where $\alpha_{\mathrm{exp}}$ is an experimental calibration constant. $A(t)$ is proportional to the spin polarization of the muon ensemble.

The asymmetry $A(t)$ was initially fitted to a model with three components (two oscillatory and one non-oscillatory) across all temperatures:
\begin{equation}
  A(t) = A_{\mathrm{rel}} \left[p_1 e^{-\lambda_1t}\cos\left(2\pi \alpha f t\right)+ \right. p_2 e^{-\lambda_2t}\cos\left(2\pi f t\right)\left.+(1-p_1-p_2)e^{-\lambda_3t}\right]+A_\mathrm{bg},\label{eqn:asymmetry-fit-1}
\end{equation}
where $A_\mathrm{rel}$ corresponds to the total relaxing amplitude, $p_1$ and $p_2$ are the weights of the oscillatory components and $\lambda_1$ and $\lambda_2$ are the respective relaxation rates. The constant term $A_\mathrm{bg}$ accounts for the non-relaxing contribution from those muons that stop at the sample holder/cryostat tail. Throughout the fitting procedure, $p_1$, $p_2$ and the ratio between the oscillatory frequencies $\alpha$ were fixed at 0.66, 0.10 and 0.29 respectively. Fitted values of $\lambda_1$ and $f$ are plotted against temperature in the top left panel of Figure~\ref{neut_mu_SI}. A sharp change is observed in both $\lambda_1$ and $f$ at around $T=3.2$~K, strongly indicative of a magnetic phase transition.

For $T<3.2$~K, the oscillatory components are therefore characteristic of quasi-static long-range magnetic ordering, with the two frequencies corresponding to two, magnetically distinct, muon stopping sites. The frequencies fall as the temperature  nears the transition at $T_\mathrm{N}$~\cite{tom-prb-2006}. However, unlike the usual case where the frequency vanishes above $T_\mathrm{N}$, in \Nipyz\ the oscillation in $A(t)$ persists up to $T=20$~K. Therefore, although Model I describes the data at all measured temperatures, this behaviour of the parameters revealed  at $T>3.2$~K suggests that the model is not applicable in this regime.

A second model was therefore  required to explain the $T>3.2$~K behaviour in \Nipyz. Rather than the muons being subject to a quasi-static magnetic field, here it is probable that the persistent oscillations in asymmetry arises from a muon-fluorine entanglement state, often observed in this class of materials in the disordered regime~\cite{tom}. This occurs in fluorine-containing materials when the disordered electronic spins fluctuate rapidly on the muon timescale, motionally narrowing them from the spectrum. The muon is then relaxed by its dipolar interaction with fluorine nuclear spins \cite{tom}.  Thus, in the temperature range where $\lambda_1$ and $f$ are found to be constant using Model I, the asymmetry is better fitted to the function 
\begin{equation}
A(t)=A_\mathrm{rel}\left[p_1 e^{-\lambda_\mathrm{F-\mu}t} D_z(\omega_\mathrm{F-\mu},t) +p_2e^{-\Lambda t}\right]+A_\mathrm{bg},\label{eqn:asymmetry-fit-2}
\end{equation}
where $p_1$ and $p_2$ were fixed  0.43 and 0.57 respectively. The function describing the muon-fluorine dipolar interaction $D_z(\omega_{\mathrm{F-}\mu},t)$~\cite{tom}, models the interaction of the muon with a single $I=1/2$ fluorine nucleus. Again, the  component $A_\mathrm{bg}$ accounts for the relaxing contribution from the muons stopping at the sample holder/cryostat tail. The extracted relaxation rates $\lambda_\mathrm{F-\mu}$ and frequency $\omega_\mathrm{F-\mu}$ are shown in the bottom left panel of Figure~\ref{neut_mu_SI}. We see that  $\lambda_\mathrm{F-\mu}$  diverges as temperature approaches the magnetic transition, as often observed in similar systems~\cite{steele} and providing additional evidence for our interpretation. The value of $\omega$ corresponds to a muon site with a F-$\mu^{+}$ separation of 0.11~nm, which is also typical of muon stopping sites in this class of material \cite{steele}. 

\subsection{Neutron diffraction}

\begin{figure}[t]
\centering
\includegraphics[height=6cm]{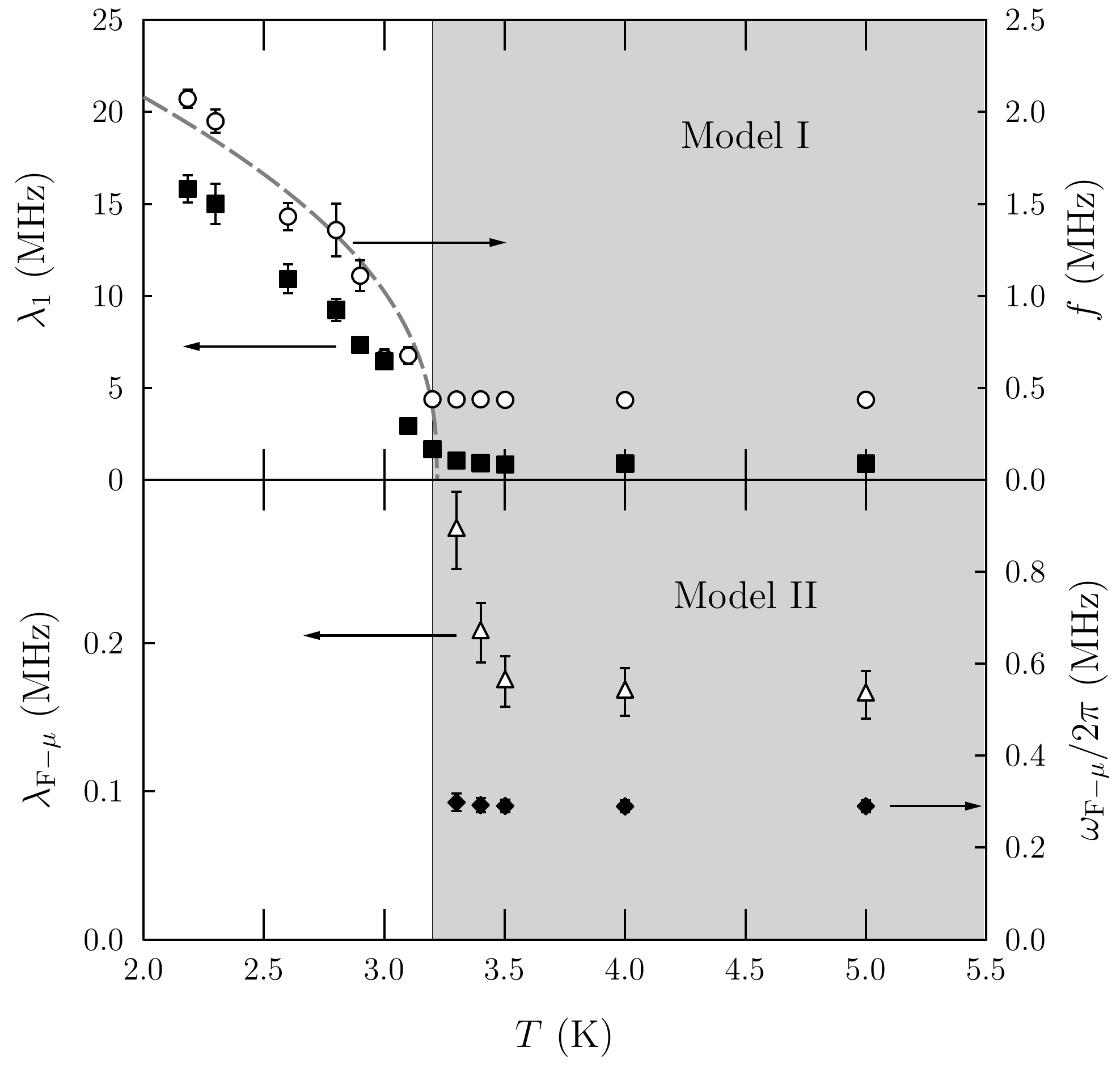}
\includegraphics[height=6cm]{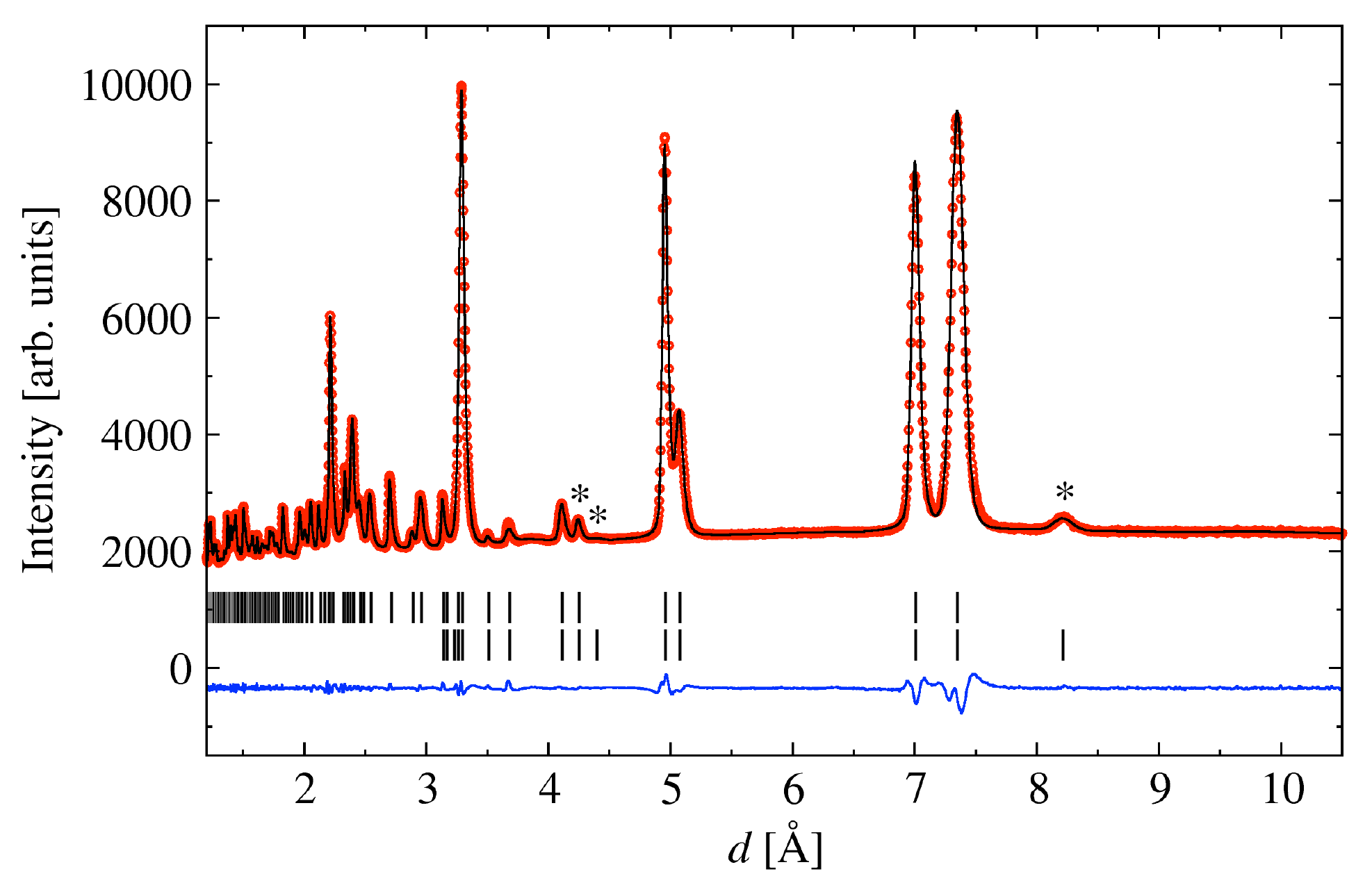}
\caption{Left: Temperature dependence of the fitted muon-spin relaxation parameters using Model I (top) and Model II, see text for details. The border of the white/grey area is at $T=3.2$~K and the dashed line is a guide to the eye. Right: Scattered neutron intensity (circles) at 1.5~K as a function of $d$-spacing for \Nipyz. The three magnetic peaks are marked with * and correspond to (from right to left) the $\{1,0,1\}$, $\{1,0,3\}$, and $\{2,1,1\}$ families of reciprocal lattice vectors. The black line is a fit to the nuclear and magnetic peaks with the Ni(II) moments lying perpendicular to the $c$-axis. Bragg peaks are indicated by ticks and the blue line is the difference between the data and the fit.} 
\label{neut_mu_SI}
\end{figure}

Magnetic diffraction patterns were recorded on the WISH diffractometer (ISIS, Rutherford Appleton Laboratory, UK)~\cite{WISH}. A deuterated polycrystalline sample, [Ni(D$_2$O)(d$_4$-pyz)$_2$]($^{11}$BF$_4$)$_2$ was loaded into a cylindrical vanadium can and placed in an Oxford Instruments cryostat with a base temperature of 1.5~K. Diffraction data were collected over the temperature interval 1.5---10~K, with long counting times (3~hrs) at 1.5~K and 10~K. Intermediate temperature points were measured with an exposure time of 0.5~hrs. Scattered neutron intensity (circles) at 1.5~K as a function of $d$-spacing fits shown in the right-hand panel of Figure~\ref{neut_mu_SI}. 

\section{Calculations}

\subsection{$S=1$ single-ion anisotropy Hamiltonian}

In an applied magnetic field ${\bf B} = \mu_0 {\bf H}$ the exchange-free $S=1$ Hamiltonian (Equation 1, main text) can be written 
\begin{equation}
\hat{\mathcal{H}}= \begin{pmatrix}
D+h_z & \frac{1}{\sqrt{2}}(h_x-ih_y) & E \\
\frac{1}{\sqrt{2}}(h_x+ih_y) & 0 & \frac{1}{\sqrt{2}}(h_x-ih_y) \\
E & \frac{1}{\sqrt{2}}(h_x+ih_y) & D-h_z \\
\end{pmatrix},
\label{ham_mat}
\end{equation}
where $h_i=g_i\mu_{\rm B}B_i$. The resulting energy eigenvalues $\epsilon_n$ are used to construct a partition function and obtain thermodynamic quantities such as magnetization, magnetic susceptibility and heat capacity. For fields along $x$, $y$ and $z$ the eigenvalues are
\begin{equation}
\begin{aligned}
\epsilon_x &= D - E &{~~~~{\rm and} ~~~~}  &\left(\frac{D+E}{2}\right)\pm\sqrt{\left(\frac{D+E}{2}\right)^2+(g_x\mu_{\rm B}B_x)^2},\\
\epsilon_y &= D + E &{~~~~{\rm and} ~~~~}  &\left(\frac{D-E}{2}\right)\pm\sqrt{\left(\frac{D-E}{2}\right)^2+(g_y\mu_{\rm B}B_y)^2},\\
\epsilon_z &= 0 &{~~~~{\rm and} ~~~~} &D\pm\sqrt{E^2+(g_z\mu_{\rm B}B_z)^2}.
\end{aligned} 
\end{equation}
In zero field these reduce to $\epsilon = D-E$, 0, and $D+E$.

\subsection{Simulations of magnetization and heat capacity}

Simulations of the powder average magnetization of an ensemble of $S = 1$ spins must consider the field evolution of the three $\epsilon_n$ eigenvalues for all angles of the applied field. For a given field strength, $H$, and orientation of the field with respect to the $z$-axis expressed in terms of the polar angles $\theta_i$, $\phi_j$, the eigenvalues of Equation~\ref{ham_mat} are deduced by diagonalization. Inserting these eigenvalues into a partition function at a fixed temperature, the magnetization at a particular field strength and orientation $M(H, \theta_i, \phi_j)$, can be deduced (see e.g.~\cite{Blundells_book}).
The polar angles are incremented (in 20 evenly spaced steps of $\Delta\theta$ and $\Delta\phi$) such that, in total, 400 orientations of the field are considered. The average magnetization, $\langle M(H)\rangle$, at a particular applied field is approximated by:
\begin{equation}
\langle M(H)\rangle = \frac{\sum_{i,j}M(H, \theta_i, \phi_j)\sin\theta_i\Delta\theta\Delta\phi}{\sum_{i,j}\sin \theta_i \Delta \theta \Delta \phi}.
\end{equation}
The resulting data are shown in Fig.~3(a) \& (c) of the main text along with ${\rm d}M/{\rm d} H$ and ${\rm d}^2M/{\rm d} H^2$. By using different temperatures in the partition function for this calculation, the temperature dependence is also explored [Fig.~3(b) \& (d), main text].

In analogous manner the heat capacity at a particular temperature and field orientation, $C_{\rm mag}(T, \theta_i, \phi_j)$  was determined from the eigenvalues of Equation~\ref{ham_mat}. The direction of the applied field was incremented and $C_{\rm mag}$ recalculated. By considering 400 different field orientations with respect to the $z$-axis, the average $\langle C_{\rm mag}(T)\rangle$ at a given field and temperature is estimated from
\begin{equation}
\langle C_{\rm mag}(T)\rangle = \frac{\sum_{i,j}C_{\rm mag}(T, \theta_i, \phi_j)\sin\theta_i\Delta\theta\Delta\phi}{\sum_{i,j}\sin \theta_i \Delta \theta \Delta \phi}.
\end{equation}
The field dependence of the heat capacity is examined by repeating the calculation at different fixed values of $H$. The results for the easy-axis and easy-plane cases with $E=0$ are shown in Fig.~4 of the main text.

\subsection{Electronic structure calculations}

\begin{figure}[t]
\centering
\includegraphics[height=6cm]{spin0011.pdf}
\includegraphics[height=6cm]{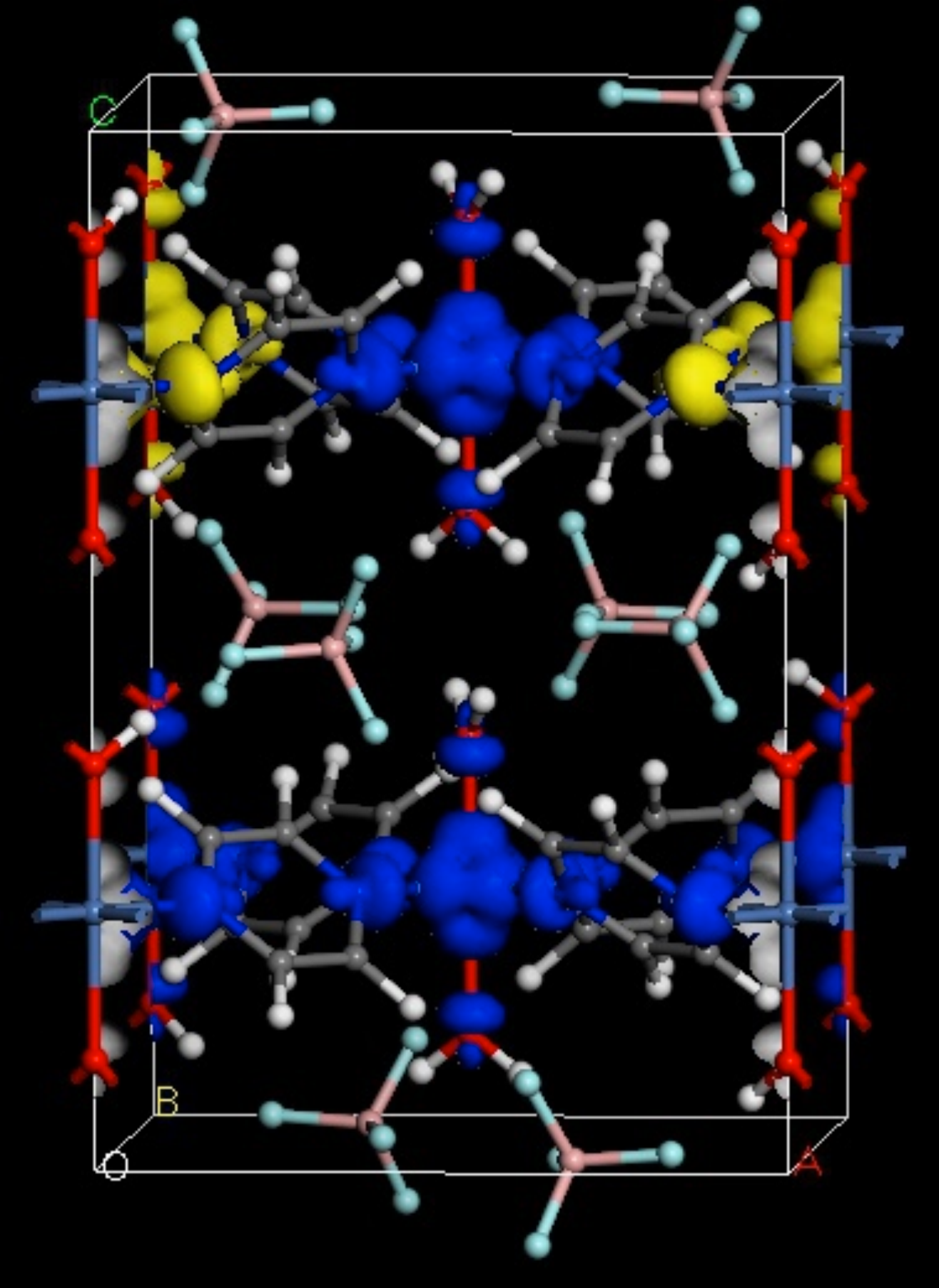}
\includegraphics[height=6cm]{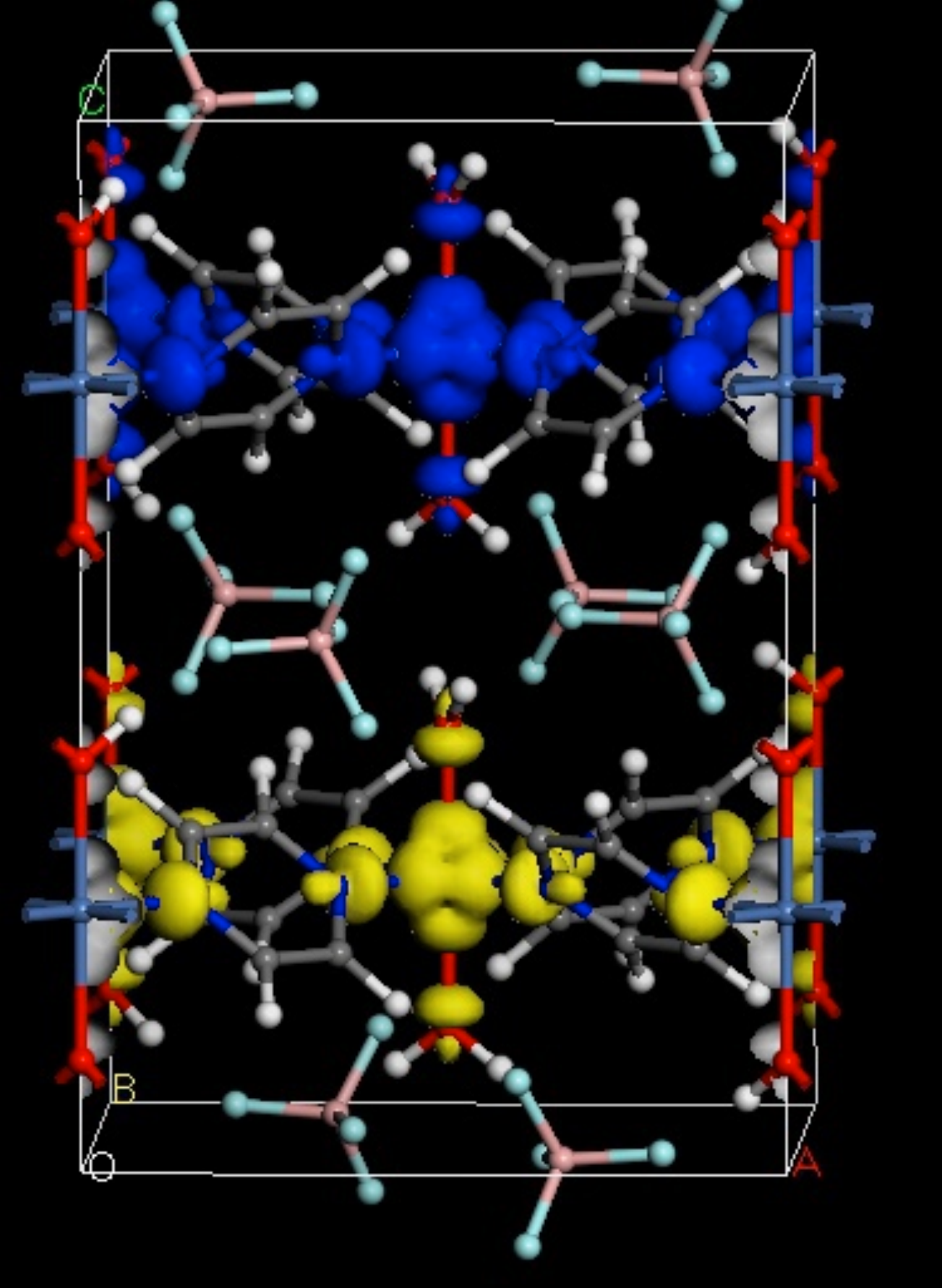}
\caption{\label{dft} Left: The lowest energy spin state; blue shows
the spin density of one spin channel, yellow the other. Middle: Spin flip within one $x-y$ plane with
respect to lowest energy spin configuration. Right: The third lowest energy configuration is
shown, which is antiferromagnetically stacked ferromagnetic planes.}
\end{figure}

Our computational approach
involves determining the energy differences between ordered spin
states that differ by a number of reversed spins\cite{JNoise,datta}.  An
underlying physical assumption, therefore, is that upon magnetically
ordering, the magnetic structure is constrained to be collinear. In
other words, the value of the magnetic exchange derived assumes that
nearest neighbor spins $\boldsymbol{S}$ at sites $i$ and $j$ obey $\boldsymbol{S}_{i}\cdot \boldsymbol{S}_{j}= \pm 1$. If this is not the
case, then the error in this assumption is absorbed into the value of
the exchange constant  that is derived. 
More specifically, we map the magnetic
centres of the system to an Ising Hamiltonian
\begin{equation}
\hat{H}  =  \sum_{ij}I_{ij}S_z^iS_z^j,
\end{equation}
where the sum is over neighbours and $I$ is the Ising exchange energy.
In \Nipyz\ the spins are (nominally)
located on the Ni atoms. There are two different pairs of Ni-Ni
interactions. The first is within $x-y$ planes with Ni-Ni distance of
$7.012$~\AA\ (coupling $I_{1}$) and the second aligned along $z$ where Ni atoms are
separated by $7.436$~\AA\ (coupling $I_{2}$).
Each Ni interacts with four neighbours with strength $I_1$ within the $x-y$ plane, and two neighbours along $z$ with
strength $I_2$. (Here we count the couplings per Ni---Ni bond, not per Ni
atom.) Therefore the $I$-coupling model simplifies to a nearest $x-y$
neighbour and nearest $z$ neighbour energy:
\begin{equation}
\hat{H} = I_{1} \sum_{\langle ij \rangle_{xy}} S_{i}S_{j}
+ I_{2} \sum_{\langle ij \rangle_{z}} S_{i}S_{j}.
\end{equation}
The spin structure of the lowest energy state is shown
in the left panel of Figure \ref{dft}. 
It forms an antiferromagnetic state within each of the $x-y$ planes
and offset by $(1/2,1/2)$ in going from the $z=1/4$ to $z=3/4$ plane.

To evaluate the intraplane $I_1$ coupling and the interplane (next
nearest neighbour) $I_2$ coupling, differences in energy of the spin
configurations are considered. Details are as follows: the second lowest
energy state is shown in the middle panel of Fig.~\ref{dft} which is 15.49~meV
higher in energy than the ground state. A single spin is flipped on
the central Ni atom in one plane which means that there is an $I_1$
contribution associated with each of the four nearest neighbours in the
$x-y$ plane. Additionally this atom has two $I_2$ neighbours along
$z$, hence
\begin{equation}
\label{J42}
2S\left(4I_1+2I_2\right)=15.49\ {\rm meV},
\end{equation}
where $S=1$ for these Ni spins. 
The next highest energy state is shown in the right-hand panel of Fig.~\ref{dft}, which is
20.52~meV above the ground state. Comparing what is flipped, the
central atom on both of the $x-y$ planes have changed. Each of these
atoms have 4 $I_1$ ($x-y$ plane) neighbours. However since both atoms
have flipped then the $I_2$ interaction ($z$-direction) is unchanged,
it is still antiferromagnetic. Therefore we have
\begin{equation}
\label{J80}
2S(8I_1+0I_2)=20.52\ {\rm meV}.
\end{equation}
Solving equations (\ref{J42}) and (\ref{J80}) gives
$I_1=1.283$~meV and  $I_2=1.308$~meV.

In order to convert these Ising couplings to Heisenberg couplings, we
employ the method from Ref.~\cite{datta}, where the estimated Heisenberg
coupling ${J}$ is related to the Ising coupling $I$ via
${J} = \frac{I}{2S}$,
where $2S$ is the spin flip factor again, that we saw above.   We conclude that the Heisenberg couplings are
${J}_{1} = 0.642$~meV
and
${J}_{2} = 0.652$~meV,
so the exchange is fairly isotropic, with magnitude around
${J}\approx 8$~K.

\section*{References}
\bibliography{NiPowderRef_SI}